\documentclass[a4paper,11pt]{article}
\usepackage[utf8]{inputenc}
\usepackage{mathtools}
\usepackage{amsmath}
\usepackage[affil-it]{authblk}
\usepackage{float}
\usepackage{subcaption}
\usepackage{verbatim}
\usepackage{enumitem}
\usepackage{authblk}

\usepackage{natbib}
\usepackage{breqn}

\usepackage{titlesec}
\usepackage[titletoc,toc,title]{appendix}
\titlespacing*{\section}{0pt}{25pt}{17pt}
\titlespacing*{\subsection}{0pt}{15pt}{12pt}

\usepackage{array} 
\newcolumntype{H}{>{\setbox0=\hbox\bgroup}c<{\egroup}@{}} 

\usepackage[bottom]{footmisc} 

\usepackage{pdflscape}

\usepackage{hyperref}
\hypersetup{colorlinks=true,urlcolor=blue,citecolor=blue,linkcolor=blue}

\usepackage{graphicx}
\graphicspath{{figures/}}

\makeatletter
\g@addto@macro\normalsize{%
  \setlength\abovedisplayskip{10pt}
  \setlength\belowdisplayskip{10pt}
  \setlength\abovedisplayshortskip{10pt}
  \setlength\belowdisplayshortskip{10pt}
}
\makeatother

\title{Regime-Switching Density Forecasts Using Economists' Scenarios}
\author{Graziano Moramarco\thanks{Department of Economics, Piazza Scaravilli 2, 40126 Bologna, Italy. 
E-mail: graziano.moramarco@unibo.it.
}}
\affil{University of Bologna}
\date{}

\begin{document}

\maketitle

    \begin{abstract}
    
		We propose an approach for generating macroeconomic density forecasts that incorporate information on multiple scenarios defined by experts. 
		We adopt a regime-switching framework in which sets of scenarios (``views") are used as Bayesian priors on economic regimes.
		Predictive densities coming from different views are then combined by optimizing objective functions of density forecasting. 
		We illustrate the approach with an empirical application to quarterly real-time forecasts of U.S. GDP growth, in which we exploit the Fed's macroeconomic scenarios used for bank stress tests. 
		We show that the approach achieves good accuracy in terms of average predictive scores and good calibration of forecast distributions. 
		Moreover, it can be used to evaluate the contribution of economists' scenarios to density forecast performance.  
		
        \vspace{0.5cm}


    \end{abstract}

    \section{Introduction}
    
    In recent years, it has become increasingly important for economic agents and policymakers to take account of the uncertainty around macroeconomic outlooks.
    In particular, two distinct approaches to this issue have emerged.
    The first one consists in generating economic forecasts in the form of (continuous) probability distributions, or density forecasts. This is now common practice among forecasters (\citealt{elliottandtimmermann2016}), and 
    many economic agents, such as financial institutions, routinely evaluate their potential losses as random draws from predictive density functions (e.g., for the computation of value at risk and expected shortfall, see \citealt{jorion2006}).
    The second approach is to define a small number of (discrete) macroeconomic scenarios (e.g., see \citealt{moodys2018}).
    This approach facilitates communication to the public regarding economic uncertainty and finds important applications in financial supervision and risk management, most notably in the design of bank stress tests, which are now integral part of the financial regulatory framework in advanced economies (e.g., \citealt{fed2018}). 
    
    This paper develops a novel forecasting approach that combines these two perspectives. 
    Using a regime-switching model, we construct macroeconomic density forecasts that explicitly incorporate information on discrete scenarios defined by economists.
    We consider alternative sets of scenarios, which we label as \textit{views}, and use them to define Bayesian priors on economic regimes in the model. 
    Next, as different views result in different density forecasts, we combine these forecasts in a way that optimizes standard evaluation criteria of density forecasting. 
    The approach is illustrated through an empirical application to U.S. GDP growth forecasts, in which we exploit the information contained in the macroeconomic scenarios defined by the Federal Reserve for its bank stress tests. 
    We show that the approach achieves good forecast accuracy and good calibration of forecast distributions.
    Moreover, we find that its forecast performance is further improved when we combine it with a beta transformation of density forecasts (\citealt{GneitingRanjan2010,GneitingRanjan2013}). 
    
    On the one hand, the proposed approach is able to produce highly flexible predictive distributions.
    	Such flexibility results both from the regime-switching framework and from the forecast combination procedure.
    	Density forecasts from regime-switching models are
    	mixture distributions, by construction: they are weighted averages of regime-specific densities, where the weights consist in the probabilities of the economy ending up in the various regimes. 
    	Likewise, density forecast combinations also produce mixture distributions. 
    	 Thus, the approach can easily accommodate a wide range of departures from normality, as mixture distributions are in general non-normal even if their individual components are.
    	 This is an important property when it comes to forecasting macroeconomic variables, whose empirical distributions often deviate from Gaussianity.
    	 In particular, a number of studies, 
    	 including \citet{fagioloetal2008}, \citet{curdiaetal2014}, and \citet{ascari_fagiolo_roventini_2015},
    	 have documented a non-normal distribution of the GDP growth rate in the U.S. and other advanced economies, 
    	 mainly as a result of large downturns.
    	 \citet{acemogluetal2017} find similar results and propose a theoretical model in which systematic departures of output from the normal distribution are explained by the interaction of idiosyncratic microeconomic shocks and sectoral heterogeneity.
    	 Compared to econometric approaches that impose non-normal distributions of errors (e.g., \citealt{hansen1994}), regime-switching models allow for a more transparent economic explanation of non-normality, based on the transition between different regimes.

    On the other hand, regime-switching models appear as a natural framework for dealing with discrete scenarios such as those used in bank stress tests. First of all, these scenarios are typically constructed in a way that directly relates to economic regimes.
    For instance, the Fed defines its scenarios as ``sets of economic and financial conditions", with baseline scenarios reflecting the most likely conditions and adverse/severely adverse scenarios reflecting conditions that prevail in recessions (\citealt{fed2017}, Appendix A of Part 252).\footnote{
    	As explained in \citet{fed2017} (Appendix A of Part 252), ``the baseline scenario is designed to represent a consensus expectation" and is constructed using the forecasts of private sector forecasters (e.g., Blue Chip Consensus Forecasts and the Survey of Professional Forecasters), government agencies, and other public-sector organizations (e.g., the International Monetary Fund and the Organization for Economic Co-operation and Development).
    	To develop the severely adverse scenario, the Fed relies on a recession approach, i.e., it specifies the ``future paths of variables to reflect conditions that characterize post-war U.S. recessions". 
    	The severity of recessions is mainly determined by the level and increase of the unemployment rate, as well as the occurrence of a housing recession.
    	In post-war U.S. recessions that are classified as severe, GDP has dropped about 3.5 percent and unemployment has increased by a total of about 4 percentage points, on average (\citealt{fed2017}, Appendix A of Part 252, Table 1).
    	The adverse scenario reflects ``a set of economic and financial conditions that are more adverse than those associated with the baseline scenario but less severe than those associated with the severely adverse scenario".
	}
    Also, many contributions in the literature on stress testing 
    call for an important role of regime-switching models in the design of macro scenarios, arguing that realistic scenarios should reflect the non-linearities that derive from the state dependency of macro-financial conditions (\citealt{adrianIMF2020}, \citealt{hanleika2019}, \citealt{borio2014}, \citealt{gross_henry_rancoita_2022}, \citealt{BidderMcKenna2015}). 
    As highlighted by \citet{hamilton2016}, Markov-switching models provide a sufficiently parsimonious and yet robust characterization of the transition into and out of recession regimes, at least for the U.S. economy.

     This paper contributes to several strands of the forecasting literature.
     First, it relates to other papers that incorporate external --possibly judgmental-- forecasts into econometric model-based forecasts. 
     \citet{Kruegeretal2017} combine forecasts from Bayesian VARs with forecasts from other sources using entropic tilting, a method for modifying a distribution in a way that satisfies specific moment conditions (\citealt{Robertsonetal2005}). 
     \citet{FaustWright2009} use the Federal Reserve Board's Greenbook forecasts as data in autoregressive and factor-augmented autoregressive models, in order to produce forecasts of GDP growth and inflation. \citet{SchorfheideSong1025} and \citet{Wolters2015} also use nowcasts from the Fed's Greenbook as data to generate forecasts from a Bayesian VAR (BVAR) and DSGE models, respectively.
     The main difference between our approach and these approaches is that we exploit information on multiple economic scenarios from external sources, mapping them into different regimes, 
     while these approaches only target point forecasts or forecast variance.

      Second, the paper relates to other contributions on forecasting using regime-switching models. 
      The most closely related approaches are Bayesian ones, such as those by \citet{pesaranetal2006}, who use a break point model (a generalization of regime-switching models) with hyperparameter uncertainty, and by \citet{bauwensetal2017}, who estimate a Markov-switching model with an unknown and potentially infinite number of regimes. 
     Unlike these studies, our paper exploits experts' views to define priors on economic regimes.
     A broad literature has investigated the forecast performance of regime-switching models.   
     While the available evidence on the accuracy of point forecasts is mixed (\citealt{elliottandtimmermann2016}), 
     these models have proved very useful for density forecasting and forecasts of tail events. 
     \citet{chauvetpotter2013} show that Markov switching models can achieve high accuracy in forecasting GDP, especially with respect to the timing and depth of recessions.  
     \citet{gewekeandamisano2011} show the usefulness of Markov mixtures for density forecasts of the stock market. 
     \citet{bauwensetal2017} use an infinite Markov-switching autoregressive moving average model to produce density forecasts of GDP.
     \citet{alessandrimumtaz2017} find that a Threshold VAR in which regime shifts depend on financial conditions produces good density forecasts of U.S. GDP during the Great Recession.

   	Third, the paper relates to the large literature on forecast combinations.
     In particular, it has been shown that gains in density forecast performance are often achieved by combining different predictive distributions. 
     Contributions in the field of macroeconomic forecasting include \citet{hallandmitchell2007}, \citet{gewekeandamisano2011}, and \citet{ganics2017}, among others. 
    A vast body of literature in other fields, 
    such as management science, risk analysis, and meteorology, also deals with the combination of probabilistic forecasts from experts into ``consensus" distributions (e.g., \citealt{GenestZidek1986}; \citealt{clemenwinkler1999}).
    For comprehensive reviews, see \citet{aastveitetal2018} and \citet{wangetal2022}.
    Several ways to further enhance density forecasts of non-normal variables have been proposed in the literature on forecast combinations, such as beta opinion pools (\citealt{GneitingRanjan2010,GneitingRanjan2013}) and empirically-transformed opinion pools (\citealt{garrattetal2023}), which will be explored in the empirical part of the paper.

    We combine forecasts from different experts' views using two main criteria of density forecast evaluation. 
     The first one is the sum of log scores, which measures the ability to assign high probabilities to outcomes that are truly likely to be observed. The second is a uniformity test on the probability integral transforms (PITs) of the forecasts, which measures the degree of calibration of the forecast distribution (\citealt{dieboldetal1998}).\footnote{A well-calibrated forecast is one that does not make systematic errors: if \textit{p} is the predicted probability assigned to a given random event, then that event should empirically occur with frequency \textit{p}.} 
     Both measures have been used in the literature to compute ``optimal" forecast combinations (\citealt{hallandmitchell2007}; \citealt{gewekeandamisano2011}; \citealt{ganics2017}).

    The remainder of the paper is organized as follows: 
    Section \ref{sec:methodology} explains the methodology, 
    Section \ref{sec:application} presents the empirical application, 
    and
    Section \ref{sec:conclusions} concludes.

	\section{Methodology}\label{sec:methodology}

		\subsection{The regime-switching model}

		Our reference model is a Markov-switching autoregressive (MSAR) model in which the intercept and the error variance depend on the unobserved state of the economy. Let $y_t$ denote a macroeconomic variable of interest at time $t$. The MSAR can be written as:
			\begin{gather}\label{eq:MSAR}
				y_t= \sum_{j=1}^{p}\alpha_j y_{t-j} + \beta_{S_t}  + \varepsilon_t \\
				\varepsilon_t \sim N(0,\sigma^2_{S_t}) \nonumber
			\end{gather}
			where $S_t$ denotes the unobserved state variable at time $t$, $\beta_{S_t}$ is the intercept in regime $S_t$, $\alpha_j$ for $j=1,\dots,p$ is a state-independent autoregressive coefficient, $p$ is the maximum lag, $\varepsilon_t$ is the error term and $\sigma^2_{S_t}$ is the regime-dependent variance of the error.\footnote{We assume state-independent autoregressive coefficients (see, e.g., \citealt{hamilton1989}) to avoid the overfitting problems that easily arise in macro applications of regime-switching models. In particular, as explained in \citet{hamilton2016}, ``inference about parameters that only show up in regime $i$ can only come from observations within that regime. With postwar quarterly data that would mean about 50 observations from which to estimate all the parameters operating during recessions. One or two parameters could be estimated fairly well, but overfitting is again a potential concern in models with many parameters. For this reason researchers may want to limit the focus to a few of the most important parameters that are likely to change, such as the intercept and the variance".
				
			Also, as pointed out by \citet{hamilton2016}, while there is no theoretical reason to assume that all regime-specific densities are normal,  
			the same concerns of overfitting suggest using a normal distribution, which is defined by just two parameters.
			The assumption of normality is generally made in the literature on regime-switching models, with very few exceptions, such as \citet{dueker1997}.
			Some empirical evidence may be used to corroborate this assumption. For instance, \citet{acemogluetal2017} find that, once large downturns are excluded from the data, the U.S. GDP growth distribution appears to be well approximated by a normal.
		} In particular, $S_t$ is a Markov chain characterized by a transition matrix $\boldsymbol{\xi}$, where the element $\xi_{kj}$ in row $k$ and column $j$ represents the probability of transition from state $k$ to state $j$:
		\begin{equation}\label{eq:transition_probability}
				\xi_{kj} = Pr (S_t=j|S_{t-1} =k)
		\end{equation}
		with $k,j=1,\dots,K$, where $K$ is the number of regimes in the economy.
		 Accordingly, the MSAR captures the typical autocorrelation of macro variables in two ways: by means of the autoregressive coefficients in \eqref{eq:MSAR} and through the persistence in the state variable $S_t$, as expressed by the transition matrix.
		Finally, let $\boldsymbol{\vartheta}$ denote the vector of parameters of the MSAR model, i.e.
			$\boldsymbol{\vartheta} = (\beta_1,\dots,\beta_K,\sigma_1,\dots,\sigma_K,\alpha_1,\dots,\alpha_p,\boldsymbol{\xi})$. 
         
         \subsection{Incorporating information on discrete scenarios}\label{subsec:theory_priors}

         The MSAR model \eqref{eq:MSAR} is estimated using Bayesian methods (see the Appendix for details).
         The priors on the parameters follow conventional distributions (\citealt{fruehwirthschnatter2006}), which are: 
         \begin{gather}
    	     \alpha_j \sim \mathcal{N}\left(a_{j,0},A_{j,0}\right)\\
         	\beta_k \sim \mathcal{N}\left(b_{0,k},B_{0,k}\right)\\
         	\sigma_{k}^2 \sim \mathcal{G}^{-1}\left(c_0,C_0\right)
         \end{gather}         
         for  $j=1,\dots,p$, where $\mathcal{N}$ and  $\mathcal{G}^{-1}$ denote Normal and inverse Gamma distributions, respectively, and $a_{j,0},A_{j,0}, b_{0,k},B_{0,k},$ $c_0,C_0$ are hyperparameters to be selected by the researcher. 
         In addition, for the transition matrix  $\boldsymbol{\xi}$ it is assumed that the rows are independent and each row follows a Dirichlet distribution, denoted by $\mathcal{D}$, i.e.: 
         \begin{equation}
         	\boldsymbol{\xi_k} \sim \mathcal{D}\left(e_{k1},\ldots,e_{kK} \right)
         \end{equation}
         where $e_{k1},\ldots,e_{kK}$ are hyperparameters, for $k=1,\dots,K$.
         
         We define a \textit{view} as a set of assumptions concerning the number of regimes $K$ and the model parameters, i.e., a specific set of values for $K$ and for the hyperparameters of the model's prior.

         Views can incorporate information on discrete economic scenarios defined by experts. 
         In particular, we consider priors in which
         each regime is ``centered" on a corresponding scenario using the following rule. 
         Consider an AR($p$) model where the coefficients are given by the $k$-state-specific prior hyperparameters, i.e.:
         \begin{equation}
         	y_t= \sum_{j=1}^{p}a_j y_{t-j} + b_{0,k}  + \varepsilon_t
         \end{equation} 
         In this model, the unconditional expectation of $y_t$, which we denote by $E_k \left( y_t \right)$, is:
         \begin{equation}\label{eq:scenario_match}
         	E_k \left( y_t \right) = \frac{b_{0,k}}{ 1- \sum_{j=1}^{p} a_j }
         \end{equation} 
         Then, given an assumption on the state-independent autoregressive coefficients $a_j$, each regime-specific hyperparameter $b_{0,k}$ is chosen in such a way that expectation \eqref{eq:scenario_match} matches a specific value derived from a scenario provided by external sources.
         The hyperparameter $B_{0,k}$ determines the tightness of the prior around these expectations or, in other words, the strength of economists' views.
         In principle, all the other hyperparameters can also be set to values suggested by economists, if available. Otherwise, uninformative or diffuse priors can be used for the remaining parameters of the MSAR. 

         When several alternative views are considered, Bayesian averaging across views can be performed.
         Let $\boldsymbol{\vartheta}_{K,i}^0$ denote the generic $i$-th view assuming $K$ states 
         and let $\boldsymbol{\pi}^{0}$ be a vector containing discrete prior probabilities assigned to all views considered.
         The posterior probability of view $\boldsymbol{\vartheta}^0_{K,i}$, which we denote as $\pi_{K,i}$, will depend on the prior probability vector $\boldsymbol{\pi}^0$ and on the values of the marginal likelihood of the MSAR model associated with the different views.
        	Note that the letter $\pi$ is used throughout the text to denote discrete probability distributions.
        	
         Further details are provided in the Appendix.

		\subsection{Optimizing density forecasts }\label{subsec:optimizing}
		
		Let us define the vector containing all observations up to time $t$ as $\mathbf{y}_t$, i.e., $\mathbf{y}_t=(y_1,\dots,y_t)$. Also, let us assume that the current time period is $T$ and the forecast horizon is one period. 
		Finally, let us denote the density forecast produced by any given view as 
		$ p\left(y_{T+1}|\mathbf{y}_{T},\boldsymbol{\vartheta}_{K,i}^0 \right)$ (see the Appendix A.2 for details on how density forecasts are calculated), 
		and let us assume that, for any given number of states $K =1, \dots, \overline{K}$, a number $P_K$ of alternative views are available.
		We consider two methods for pooling forecasts across different views:
		\begin{itemize}
			\item
			\textit{Forecast combinations}. We can express a forecast combination across MSAR views, where the vector of combination weights is denoted by $ \mathbf{w}$, as:
			\begin{equation}\label{eq:forecast_w}
				p\left(y_{T+1}|\mathbf{y}_T, \mathbf{w} \right)=\sum_{K=1}^{\overline{K}}\sum_{i=1}^{P_K}  p\left(y_{T+1}|\mathbf{y}_{T},\mathbf{\boldsymbol{\vartheta}}_{K,i}^0\right)  w_{K,i} 
			\end{equation}
			where $w_{K,i} \geq 0$ is the weight assigned to view $\boldsymbol{\vartheta}_{K,i}^0$, with $\sum_{K=1}^{\overline{K}}\sum_{i=1}^{P_K} w_{K,i} = 1$.
			\item 	
			\textit{Bayesian averaging} over different views, i.e.:
			\begin{equation}\label{eq:forecast_pi0}
				p\left(y_{T+1}|\mathbf{y}_{T},\boldsymbol{\pi}^0\right)=\sum_{K=1}^{\overline{K}}\sum_{i=1}^{P_K}p\left(y_{T+1}|\mathbf{y}_{T},\mathbf{\boldsymbol{\vartheta}}_{K,i}^0\right)\pi_{K,i}
			\end{equation}
			Forecast  \eqref{eq:forecast_pi0} is a weighted average in which the weight assigned to the view-specific forecast $p\left(y_{T+1}|\mathbf{y}_{T},\mathbf{\boldsymbol{\vartheta}}_{K,i}^0\right)$ is given by the posterior probability of view $\mathbf{\boldsymbol{\vartheta}}_{K,i}^0$, i.e., $\pi_{K,i}$, which depends on the prior probability vector $\boldsymbol{\pi}^{0}$.
		\end{itemize}
		We find the ``optimal" priors $\boldsymbol{\pi}^0$ and weights $\mathbf{w}$
		by maximizing two alternative objective functions, based on statistics that are commonly used to evaluate density forecast performance:
		\begin{enumerate}
			\item  
			\textit{Log scores}.
				The log score is the log of the predictive density function evaluated at the actual realization of the target variable. 
				Given a generic forecast horizon $h$, let $y_{t+h}^o$ (where ${}^o$ stands for ``observed") denote the realization of variable $y$ at time $t+h$, which is not observed at time $t$, when the forecast for $t+h$ is produced. 
				Also, let $R$ be the length of the timespan over which forecasts are optimized. The first objective function, denoted by $f_1$, is given by the sum of log scores over the period of interest. 
		 	Specifically, our first objective function at time $\tau$ is:
			\begin{gather}
				f_{1,\tau}\left(\boldsymbol{\omega}\right)=\sum_{t=\tau-h-R+1}^{\tau-h}\ln{\left(p\left(y_{t+h}^o|\mathbf{y}_{t},\boldsymbol{\omega}\right)\right)}
			\end{gather}
			where $\boldsymbol{\omega}$ is a place-holder for either $\mathbf{w}$ or $\boldsymbol{\pi}^0$, depending on whether we consider forecast combinations or Bayesian averaging.
			\item \textit{Probability integral transforms (PITs)}. 
			The PIT is the cumulative predictive density function evaluated at the actual realization of the variable.
			If the density forecast used to compute the PIT corresponds to the true distribution of the variable, then, for $h = 1$, the PIT values are the realizations of independently and identically distributed (i.i.d.) Uniform $(0, 1)$ variables (\citealt{dieboldetal1998}).
			Therefore, a uniformity test on the PITs can be seen as a test of correct specification of the density forecasts (see also \citealt{RossiSekhposyan2014}). Accordingly, the second objective function is given by: 
			\begin{gather}
				f_{2,\tau}\left(\boldsymbol{\omega} \right)= -ks\left(\left \{\Phi\left(y_{t+1}^o|\mathbf{y}_{t},\boldsymbol{\omega}\right) \right \}_{t=\tau-R}^{\tau-1}\right)
			\end{gather}
			where $\Phi\left(\cdot\right)$ denotes the cumulative predictive density function, 
			$ks(\cdot)$ indicates the function returning the test statistics of the Kolmogorov-Smirnov (KS) test of uniformity and, as before, $\boldsymbol{\omega}$ is a place-holder for either $\mathbf{w}$ or $\boldsymbol{\pi}^0$. Maximizing $-ks(\cdot)$ is equivalent to maximizing the p-value of the KS test (whose null hypothesis is uniformity).
		\end{enumerate}

		Both the optimization based on $f_1$ and the one based on $f_2$ are solved numerically. For each $f_i$, with $i=1,2$, the optimization algorithm delivers two vectors at time $\tau$: the vector of optimal forecast weights $ 	\mathbf{w}^\ast_{i,\tau}$ for the set of alternative views, i.e.:
		\begin{equation}\label{eq:optimal_weights}
		\mathbf{w}^\ast_{i,\tau} = \underset{\mathbf{w}}{\arg\max} \ f_{i,\tau}\left(\mathbf{w}\right)
		\end{equation}
		and the vector of optimal prior probabilities $\boldsymbol{\pi}^{0\ast}_{i,\tau}$:
		\begin{equation}\label{eq:optimal_priors}
		\boldsymbol{\pi}^{0\ast}_{i,\tau} = \underset{\boldsymbol{\pi^0}}{\arg\max} \ f_{i,\tau}\left(  \boldsymbol{\pi^0}  \right)
		\end{equation}
		The former represents the typical problem explored in the literature on density forecast combination, whereas the latter can be seen as an empirical method for eliciting priors in the context of Bayesian model averaging. The optimal prior $\boldsymbol{\pi}_{i,\tau}^{0\ast}$ represents the discrete prior probability distribution of views such that the resulting posterior $\boldsymbol{\pi}_{i,\tau}^\ast$, when used as a vector of forecast weights, maximizes the density forecast performance, based on the selected objective function.
		In practice, the main difference between \eqref{eq:optimal_weights} and \eqref{eq:optimal_priors} is that the first problem directly delivers weights for forecast combination, while in the second case the actual forecast weights are the posterior probabilities, and so will also depend on the marginal likelihoods of all views, i.e.
		$p(\mathbf{y}|\boldsymbol{\vartheta}^0_{K,i}) $ $\forall K,i$ (see the Appendix).

		\section{Empirical application: U.S. GDP growth}\label{sec:application}
		
		This section illustrates the approach through an empirical application to density forecasts of U.S. GDP growth. We use real-time quarterly data from 1948Q1 to 2019Q4\footnote{Source: U.S. Bureau of Economic Analysis, Real Gross Domestic Product (series code: GDPC1). We use the complete set of real-time data vintages provided in the Archival FRED (ALFRED) database compiled by the Federal Reserve Bank of St. Louis. 
			The first vintage was released in December 1991. For all sample windows ending before 1991Q4, we use the data contained in the first available vintage.
			As vintages are generally released at a monthly frequency, for each quarter we take the last vintage released within that quarter.
		} 
		and consider the year-on-year growth rate of real GDP (expressed in percentage points in what follows).\footnote{Our choice of using the year-on-year (YoY) rather than the quarter-on-quarter (QoQ) growth rate of GDP in the MSAR model follows the same reasoning as in \citet{bindergross103}:
			since the YoY rate is more persistent than the QoQ rate, changes in the mean and variance of the YoY rate are more likely to reflect the transition between different economic regimes compared to changes in the QoQ rate, which are more affected by noisy events specific to a given quarter. 
		} 
		Over this period, the (unconditional) distribution of GDP growth is non-normal (the Jarque-Bera test rejects normality at the 1\% level of significance) and, in particular, fat-tailed (kurtosis is close to 4), in line with the findings of prior literature (e.g., \citealt{fagioloetal2008}; \citealt{ascari_fagiolo_roventini_2015}).
	
		We set the lag length $p$ of the model to 5, in consideration of the quarterly frequency of the variable.
		As explained below in more detail,
		we first produce pseudo-out-of-sample forecasts using a recursive window scheme.
		Then, we calculate optimal weights\ $\mathbf{w}^\ast$ and priors $\boldsymbol{\pi}^{0\ast}$ over time, following the procedure from section \ref{subsec:optimizing}.
		Finally, we assess the performance of pooled forecasts on an evaluation sample, i.e., using observations of the target variable that have not been used in the optimization procedure.

			\subsection{Priors and Fed scenarios}

			We consider a total of 13 alternative views on U.S. GDP growth regimes.
			Eight views impose strongly informative priors derived from the scenarios of the Fed stress tests 2015-2018.\footnote{The scenarios are available https://www.federalreserve.gov/supervisionreg/dfast-archive.htm.} 
			The remaining five views are ``vague" views, defined by imposing uninformative priors on MSAR parameters.
			We consider different assumptions on the number of regimes $K=1,2,3,4,5$.\footnote{Our choice of the maximum number of regimes is supported by the results of \citet{bauwensetal2017}. These authors develop a model that allows for an infinite number of regimes, using a nonparametric Dirichlet process. However, when estimating the model for U.S. GDP growth (with different breaks for the mean and variance parameters), they find that the posterior probability of the number of regimes being at most 5 lies between 98\% and 100\% for the mean parameters and between 74\% and 100\% for the variance, depending on the prior used for estimation.}
			
			Let us first consider the Fed-based views. For each of the four stress tests under consideration, two views are constructed, one with $K=3$ and the other with $K=5$.
			In the view with $K=3$, one of the regimes (which may be called the ``normal times" regime), is derived from the Fed baseline scenario, another (``adverse regime") from the adverse scenario and the last one (``severely adverse regime") from the severely adverse scenario.\footnote{Although the Fed stress scenarios represent hypothetical paths and not forecasts, they are intended to be plausible even when severe. Therefore, they can legitimately be assigned predictive probabilities (see e.g., Yuen 2013) and used to form density forecasts.} In particular, we ``center" each regime on the corresponding Fed scenario using the rule described in section \ref{subsec:theory_priors}.
			Specifically, we consider an AR(5) model where the coefficients are given by the $k$-state-specific hyperparameters of the prior $\boldsymbol{\vartheta}^0_{K,i}$, so equation \eqref{eq:scenario_match} implies
			$
				E_k \left( y_t \right) = b^{(K,i)}_{0,k} / ( 1- \sum_{j=1}^{5} a_j^{(K,i)} )
			$.
			Then, after making an assumption on the state-independent $a_j^{(K,i)}$, with $j=1,\dots,5$, each regime-specific $b^{(K,i)}_{0,k}$ is chosen in such a way that this expectation  matches a specific value derived from the relevant scenario of the Fed stress test. For the normal times regime, this value is the average growth rate in the last four quarters of the baseline scenario, which is assumed to be close to the convergence value of the year-on-year growth rate in the absence of shocks.\footnote{The stress test scenarios are defined in terms of annualized quarter-on-quarter growth rates, so that averaging over the last four quarters approximates the year-on-year growth rate in the last quarter.}
			For both the adverse and the severely adverse regimes, the value to be matched is the average growth rate in the first four quarters of the corresponding scenario, as the first quarters are those when the negative shocks are assumed to occur and the growth rates are lowest. 
			
			An example may help. Let us consider the view with $K=3$ derived from the 2018 Fed stress test. The average growth rate of GDP in the last four quarters of the baseline scenario is 2.1\%, while the average growth rates in the first four quarters of the adverse and severely adverse scenarios are -2.125\% and -6.275\% respectively. 
			As customary in the related Bayesian literature (e.g., \citealt{alessandrimumtaz2017}, \citealt{fruehwirthschnatter2006}), we set the prior on the autoregressive component of the model using a parsimonious specification. In particular, we set the prior mean of the autoregressive coefficients to the (approximate) OLS estimate of an AR(1), i.e., to 0.9 for the first lag and 0 for the higher-order lags (recall, however, that we allow up to 5 lags in the posterior estimates of the model). Then, $\sum_{j=1}^{5} a_j^{(K,i)} = 0.9$. Accordingly, the prior means for the regime-specific intercepts are set to $b_{0,1} = 2.1/(1-0.9)=0.21$ for the normal times regime,  $b_{0,2} = -2.125/(1-0.9)=-0.2125$ for the adverse regime and  $b_{0,3} = -6.275/(1-0.9)=-0.6275$ for the severely adverse regime.

			The four stress test-based views with $K=5$ expand the views with $K=3$ by adding two regimes: a regime which we may call ``recovery from adverse shock", designed to match the last four quarters of the adverse scenario, and a regime of ``recovery from severely adverse shock", which matches the last four quarters of the severely adverse scenario. This is done in consideration of the fact that growth rates in the last four quarters of the Fed's adverse and severely scenarios are assumed to be higher than the baseline rates, implying a rebound of the economy after a negative shock. Of course, such regimes may be more generally interpreted as ``favorable regimes" characterized by positive shocks and not necessarily as recoveries from recessions.  
			
			In the five vague views, all priors on the intercepts are centered on 0 and have a variance of 1 percentage point, while the priors on the autoregressive coefficients are centered on 0.5 for the first lag, on 0 for the higher-order lags, and have a variance of 1. Taken jointly, these assumptions imply a large prior variance on the regime-specific means of the GDP growth rate.
			Conversely, in the Fed-based views, the priors for both $\beta$ and $\alpha$ are strongly informative, so as to ensure that the regime-specific means are tightly centered on the stress test values. In particular, both priors are assumed to have minimal variance, equal to $ 10^{-5}$. For the autoregressive coefficients $\alpha$, the prior mean is assumed to be 0.9 for the first lag and 0 for higher-order lags, as mentioned in the previous example.
       		
       		No strong assumption is made regarding the regime-switching error variance $\sigma_k^{2}$. Instead, a diffuse hierarchical prior is assumed in all views.\footnote{
       			Specifically, a Gamma hyper-prior is defined for $C_0$:
            	\begin{equation}
        		C_{0} \sim \mathcal{G}\left(g_{0},G_{0}\right)
        		\end{equation}
        		To make the prior on $\sigma_k^{2}$ diffuse, the following values are selected for the hyperparameters: $c_0=3$, $g_0=0.5$ and $G_0=0.5$. These imply that $\sigma_{k}^2$ has a prior expected value of 0.5 percentage points of GDP and a high prior variance of 1.25 percentage points (see Appendix A.3).
       		}

			Finally, the hyperparameters for the $k$-th row of the transition matrix $ \boldsymbol{\xi}$ are set in such a way that the (prior) expected probability of remaining in the same state $k$ in the next period is $ \text{E}(\xi_{kk}) = 2/3$ regardless of the number of regimes $K$, while the probability of moving to a different, specific state $j$ decreases with the number of regimes, $ \text{E}(\xi_{kj}) = 1/[3(K-1)]$ (see \citealt{fruehwirthschnatter2006}).\footnote{
			Specifically, the hyperparameters for the $k$-th row of the transition matrix $ \boldsymbol{\xi}$ are $e_{kk} = 2$ and $e_{kj}=1/(K-1)$ if $k \neq j$, $\forall k,j$.
			Given the properties of the Dirichlet distribution, $ \text{E}(\xi_{kj}) = e_{kj}/(\sum_{l=1}^{K}e_{kl})$.
			}
			
			The summary of the alternative views is provided in Table \ref{tab:views}, where views 1-5 are the vague ones while views 6-13 are those derived from the Fed stress tests 2015-2018.  Table \ref{tab:fed_scenarios} (in the Appendix) reports the GDP scenarios of the Fed stress tests.

	\begin{table}[H]
		\caption{Alternative views for the regime-switching model of U.S. GDP growth }
		\centering
			\scalebox{0.85}[0.85]{
		\begin{tabular}{|c|c|c|c|c|c|c|c|c|c|c|}
			\hline \hline
			& &  \multicolumn{9}{|c|}{hyperparameters}\\ \cline{4-11} 
			view no.& view type &$K$ & $b_0$                                 & $B_0$ & $a_0$         & $A_0$ & $e$ & $c_0$ & $g_0$ & $G_0$ \\
			\hline
			1&vague        & 1   & 0                                     & 1          & (0.5 0 0 0 0) & 1          & 2   & 3     & 0.5   & 0.5   \\
			2&vague       & 2   & (0,0)                                 & 1          & (0.5 0 0 0 0) & 1          & 2   & 3     & 0.5   & 0.5   \\
			3&vague        & 3   & (0,0,0)                               & 1          & (0.5 0 0 0 0) & 1          & 2   & 3     & 0.5   & 0.5   \\
			4&vague        & 4   & (0,0,0,0)                             & 1          & (0.5 0 0 0 0) & 1          & 2   & 3     & 0.5   & 0.5   \\
			5&vague        & 5   & (0,0,0,0,0)                           & 1          & (0.5 0 0 0 0) & 1          & 2   & 3     & 0.5   & 0.5   \\
			\hline
			6&Fed stress test       & 3   & (0.265,-0.0475,-0.4275)               & $10^{-5}$     & (0.9 0 0 0 0) & $10^{-5}$     & 2   & 3     & 0.5   & 0.5   \\
			7&Fed stress test        & 3   & (0.2275,-0.1850,-0.5675)              & $10^{-5}$     & (0.9 0 0 0 0) & $10^{-5}$     & 2   & 3     & 0.5   & 0.5   \\
			8&Fed stress test        & 3   & (0.205,-0.1950,-0.59)                 & $10^{-5}$     & (0.9 0 0 0 0) & $10^{-5}$     & 2   & 3     & 0.5   & 0.5   \\
			9&Fed stress test        & 3   & (0.21,-0.2125,-0.6275)                & $10^{-5}$     & (0.9 0 0 0 0) & $10^{-5}$     & 2   & 3     & 0.5   & 0.5   \\
			10&Fed stress test       & 5   & (0.39, 0.1975, 0.265,-0.0475,-0.4275) & $10^{-5}$     & (0.9 0 0 0 0) & $10^{-5}$     & 2   & 3     & 0.5   & 0.5   \\
			11&Fed stress test       & 5   & (0.39, 0.3, 0.2275,-0.1850,-0.5675)   & $10^{-5}$     & (0.9 0 0 0 0) & $10^{-5}$     & 2   & 3     & 0.5   & 0.5   \\
			12&Fed stress test       & 5   & (0.39, 0.3, 0.205,-0.1950,-0.59)      & $10^{-5}$     & (0.9 0 0 0 0) & $10^{-5}$     & 2   & 3     & 0.5   & 0.5   \\
			13&Fed stress test       & 5   & (0.43, 0.32, 0.21,-0.2125,-0.6275)    & $10^{-5}$     & (0.9 0 0 0 0) & $10^{-5}$     & 2   & 3     & 0.5   & 0.5   \\
			\hline \hline
		\end{tabular}
	}
		\subcaption*{\textit{Notes:} The table lists the 13 priors (\textit{views}) used to estimate the Bayesian Markov-switching autoregressive (MSAR) model considered in the empirical application. $K$ denotes the assumed number of regimes, $b_0, B_0, a_0, A_0, e, c_0, g_0$ and $G_0$ are the hyperparameters of the priors. Please refer to Section \ref{sec:methodology} for an explanation of the parameters. Views 1-5 represent diffuse priors, while views 6-13 are strongly informative priors derived from the Fed supervisory scenarios.}
		\label{tab:views}
	\end{table}

        \subsection{Optimization scheme}\label{subsec:optimization_scheme}
\begin{figure}[H]
		\centering
		\caption{Optimization scheme}
		\includegraphics[scale=0.43]{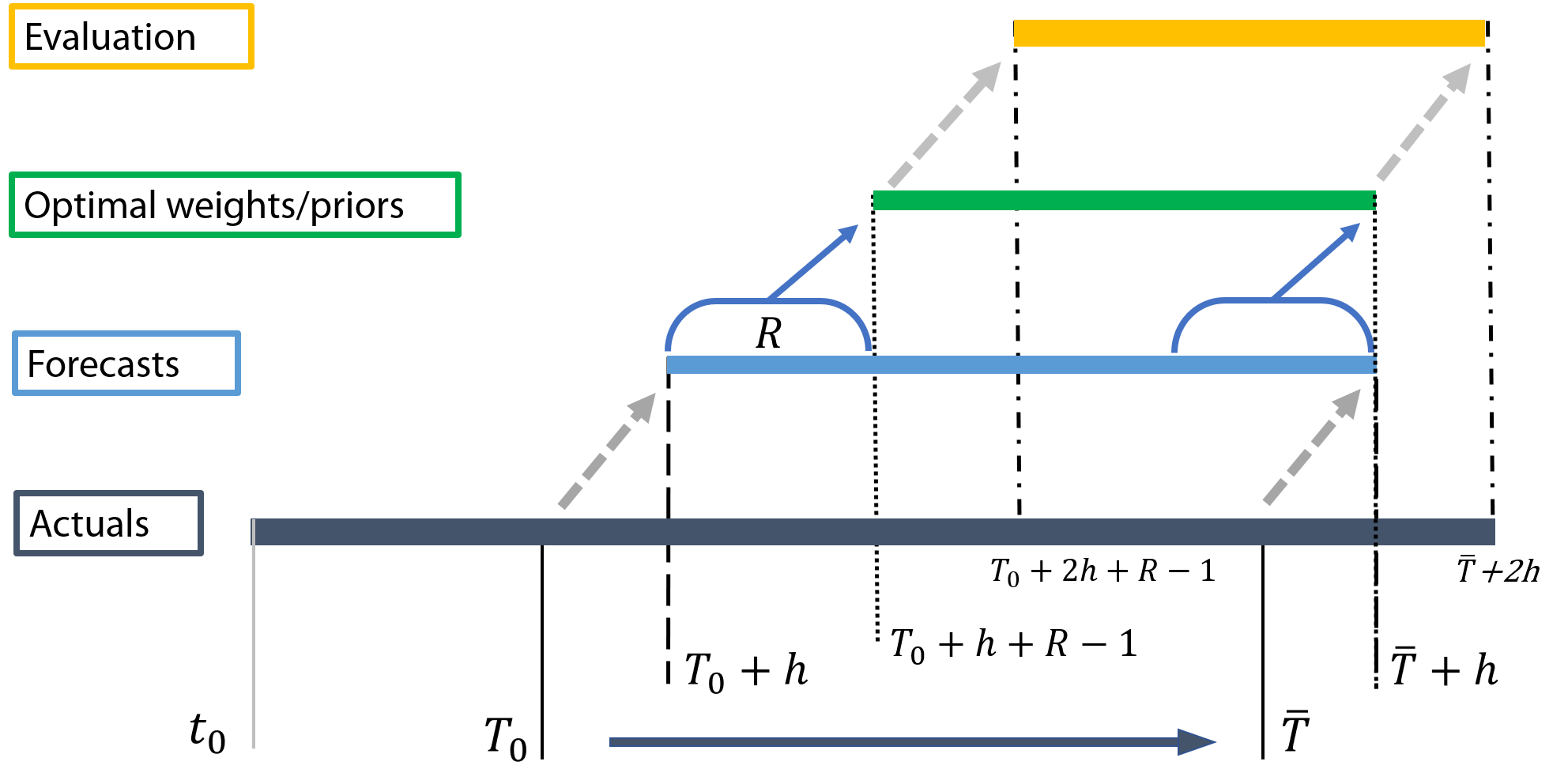}
		\subcaption*{\textit{Notes:} The figure summarizes the density forecast optimization scheme. First, the MSAR model is recursively estimated on actual GDP data (dark blue bar) using alternative views. The sample start date is denoted with $t_0$, the end date runs from $T_0$ to $\overline{T}$. For each sample window, the estimates generate density forecasts with horizon $h$ (light blue bar). A rolling sequence of $R$ forecasts is used to compute optimal forecast weights and prior probabilities (green bar) for the views. The optimal weights/priors obtained in each period are used to combine the view-specific forecasts for subsequent periods. The resulting composite forecasts (dark yellow bar) are evaluated by comparison with the actual data over the period from $T_0+2h+R-1$ to $\overline{T}+2h$.}
		\label{fig:optimization_scheme}
	\end{figure}

        We generate a sequence of pseudo-out-of-sample density forecasts using a recursive-window estimation scheme.\footnote{In this context, the choice of using expanding windows, as opposed to rolling windows, increases the probability that the variable ``visits" most or all the regimes within the sample, thus greatly facilitating the estimation of the regime-switching model.} 
        Next, the forecasts are used to carry out the optimization of weights/priors, which is iterated over time. The procedure can be described as follows. 
       Let us assume that we are at time $T_w$ and the forecast horizon is $h$.
       For each view under consideration, the MSAR model is recursively estimated using observations between an initial time $t_0$ and time $t$, with $t=T_0,T_0+1,\dots,T_w-h$. $T_0$ is therefore the end period of the shortest estimation sample. Estimates at $T_0$ are used to make forecasts for period $T_0+h$, estimates at $T_0+1$ are used to make forecasts for $T_0+1+h$, and so on.
       Thus, at time $T_w$, a sequence of past forecasts is available for each view. At this point, the algorithm computes the optimal weights/priors based on the last $R$ forecasts, i.e., maximizes the relevant objective function between $T_w-R+1$ and $T_w$. Once the optimal weights/priors are retrieved, they are used to combine the different view-specific forecasts for the future period $T_w+h$, which is out of the optimization sample. When the actual value of the variable of interest is observed, at time $T_w+h$, the performance of the combined forecast is measured. The index $T_w$ runs from $T_0+h+R-1$ to $\overline{T}+h$, where $\overline{T}$ is the end of the largest estimation sample. $\overline{T}+2h$ is the last available observation for the target variable. Therefore, the period from $T_0+2h+R-1$ to $\overline{T}+2h$ defines the evaluation sample, or test set. Figure \ref{fig:optimization_scheme} summarizes the procedure, which closely follows \citet{ganics2017}.

         More specifically, the application to U.S. GDP growth sets $t_0$=1948Q1, $T_0$=1967Q4, $R$=40 quarters, $h$=1 quarter and $\overline{T}$=2019Q2. Accordingly, the evaluation sample runs from 1978Q1 to 2019Q4.\footnote{We estimate the MSAR model using the MATLAB package \texttt{bayesf Version 2.0} by \citet{fruehwirthschnatter2008}. For each MSAR estimate, the Markov Chain Monte Carlo (MCMC) algorithm uses 1000 iterations as burn-in and 1000 iterations to store the results. Starting from the sample of forecasts produced by the MCMC algorithm, a complete probability density function is fitted using a normal kernel smoothing function.}
We focus on the forecast horizon $h=1$ because, as previously mentioned, the result of PIT uniformity of well-behaved forecasts, used the optimization, only holds for $h=1$ (see \citealt{RossiSekhposyan2014}).

        \subsection{Results}\label{subsec:results}
        
        \subsubsection{Main results}
       
        Table \ref{tab:results} shows the performance of our composite regime-switching forecasts using optimal forecast weights and optimal priors, and compares it with benchmark approaches.  
        We label our forecasts as scenario-augmented MSAR (SA-MSAR) forecasts.
        As mentioned in section \ref{subsec:optimizing}, weights $\mathbf{w}_{1}^{*}$ and priors $\boldsymbol{\pi}_{1}^{*}$ result from the optimization based on the sum of log scores, while $\mathbf{w}_{2}^{*}$ and $\boldsymbol{\pi}_{2}^{*}$ are obtained by maximizing the p-value of the Kolmogorov-Smirnov (KS) test of uniformity for the PITs.
        The first benchmark considered is a simple AR model (corresponding to view 1 in Table \ref{tab:views}). 
       	Next, we consider three models that allow for non-normal and heteroskedastic errors: an AR with Student-$t$ errors, an AR with ARCH errors and an AR with GARCH errors.
       	As with the MSAR models, the lag length for the AR component is set to 5 for all models, while the ARCH and GARCH components have a lag length of 1.
        The remaining two benchmarks consist in uniform combination schemes over the alternative views, assigning equal forecast weights/prior probabilities to different values of $K$ and, for any given $K$, equal weights/probabilities to the alternative views defined using $K$ regimes.\footnote{For instance, in the case of equal prior probabilities, it is assumed that $\pi_K^0=1/\overline{K}$ for each $K$ and that $\pi(\boldsymbol{\vartheta}_{K,i}^0|K)=1/P_K$ for each view $\boldsymbol{\vartheta}_{K,i}^0$.} 
                
        The table shows the average predictive density (APD), i.e., the average of the exponential of log scores, and the p-value of the KS test.

       The results indicate that our optimized SA-MSAR forecasts achieve good forecast accuracy and good calibration of forecast distributions.  
       When optimized by log scores, the SA-MSAR forecasts outperform all benchmarks in terms of APDs.
       When optimized by PITs, they lead to non-rejection of the null hypothesis of uniformity at the 5\% level in the KS test, indicating a reliable specification of the conditional predictive distribution of GDP growth. Thus, optimized SA-MSAR forecasts forecasts achieve well-behaved PITs.\footnote{
       	Since correct calibration implies that the PITs are realizations of i.i.d U(0,1) variables, we also test for the serial independence of the PITs. Specifically, following \citet{RossiSekhposyan2014}, we perform a Ljung–Box test of independence for both the first and the second moment of the PITs.  Both tests do not reject the null hypothesis of serial independence, implying that forecasts are well-calibrated also in this respect.} 
       By contrast, all benchmarks lead to rejection of the PIT uniformity hypothesis. 

             \begin{table}[H]
       	\caption{Density forecast performance}
       	\centering
       	{
       		\begin{tabular}{l|ll|HH}
       			\hline \hline
       			forecasting method & APD   & KS   & LB1  & LB2   \\
       			\hline
       			AR				        & 0.28 & 0.00 &  & \\
       			AR-\textit{t}  &  0.31 & 0.00 &  &  \\
       			AR-ARCH       & 0.30 & 0.00 & 0.11 & 0.04 \\
       			AR-GARCH  & 0.21  & 0.00 &  &  \\
       			SA-MSAR - Equal forecast weights   & 0.32 & 0.00 &  & \\
       			SA-MSAR - Equal prior probabilities       & 0.36  & 0.00 &  &  \\
       			\hline
       			\textbf{SA-MSAR - Optimal weights} $\mathbf{w}_{1}^*$ & \textbf{0.37} & \textbf{0.00} &  & \\
       			\textbf{SA-MSAR - Optimal priors }$\boldsymbol{\pi}_{1}^*$  &\textbf{0.37} & \textbf{0.00} &  & \\
       			\textbf{SA-MSAR - Optimal weights} $\mathbf{w}_{2}^*$ & \textbf{0.33}  & \textbf{0.07} &  & \\
       			\textbf{SA-MSAR - Optimal priors} $\boldsymbol{\pi}_{2}^*$ & \textbf{0.34}  & \textbf{0.06} &  & \\
       			\hline \hline
       		\end{tabular}
       	}
       	\subcaption*{\textit{Notes:} The table reports the density forecast performance of our scenario-augmented Markov-switching autoregressive (SA-MSAR) model for U.S. GDP using optimal pools of views, and compares it with several benchmarks. 
       	The optimal pools include log-score-based forecast combinations (optimal weights $\mathbf{w}_{1}^*$), log-score-based Bayesian averaging (optimal prior probabilities $\boldsymbol{\pi}_{1}^*$), PIT-based forecast combinations (optimal weights $\mathbf{w}_{2}^*$), and PIT-based Bayesian averaging (optimal prior probabilities $\boldsymbol{\pi}_{12}^*$). APD denotes the average predictive density, KS denotes the p-value of the Kolmogorov-Smirnov test of uniformity of the PITs. All statistics are computed over the period 1978Q1-2019Q4.
       	}
       	\label{tab:results}
       \end{table}

\begin{figure}[H]
      	\caption{Optimal log-score-based forecast combination weights ($\mathbf{w}_1^{*}$) over time}
      	\centering
      	\includegraphics[scale=0.15]{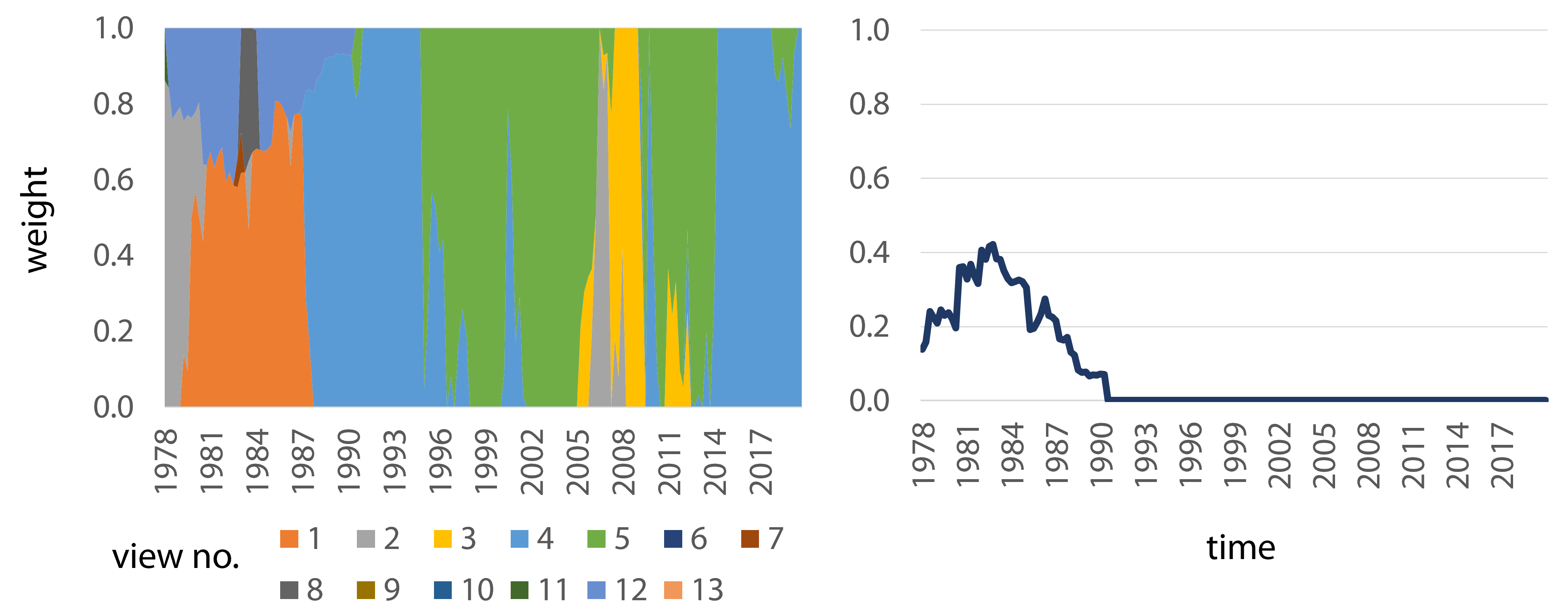}
      	\subcaption*{\textit{Notes:} The area chart in the left panel shows the time-varying forecast combination weights for all the 13 views used to estimate the Markov-switching AR model. The chart goes from 1978Q1 to 2019Q4. The weights ($\mathbf{w}_1^{*}$) are obtained using the log-score-based optimization procedure described in the paper. The right panel plots the cumulative weight assigned to the views derived from Fed supervisory scenarios (views 6-13).  See Table \ref{tab:views} for the list of views.}
      	\label{fig:w1}
      \end{figure}

\begin{figure}[H]
	  	\caption{Optimal PIT-based forecast combination weights ($\mathbf{w}_2^{*}$) over time}
	  	\centering
	  	\includegraphics[scale=0.15]{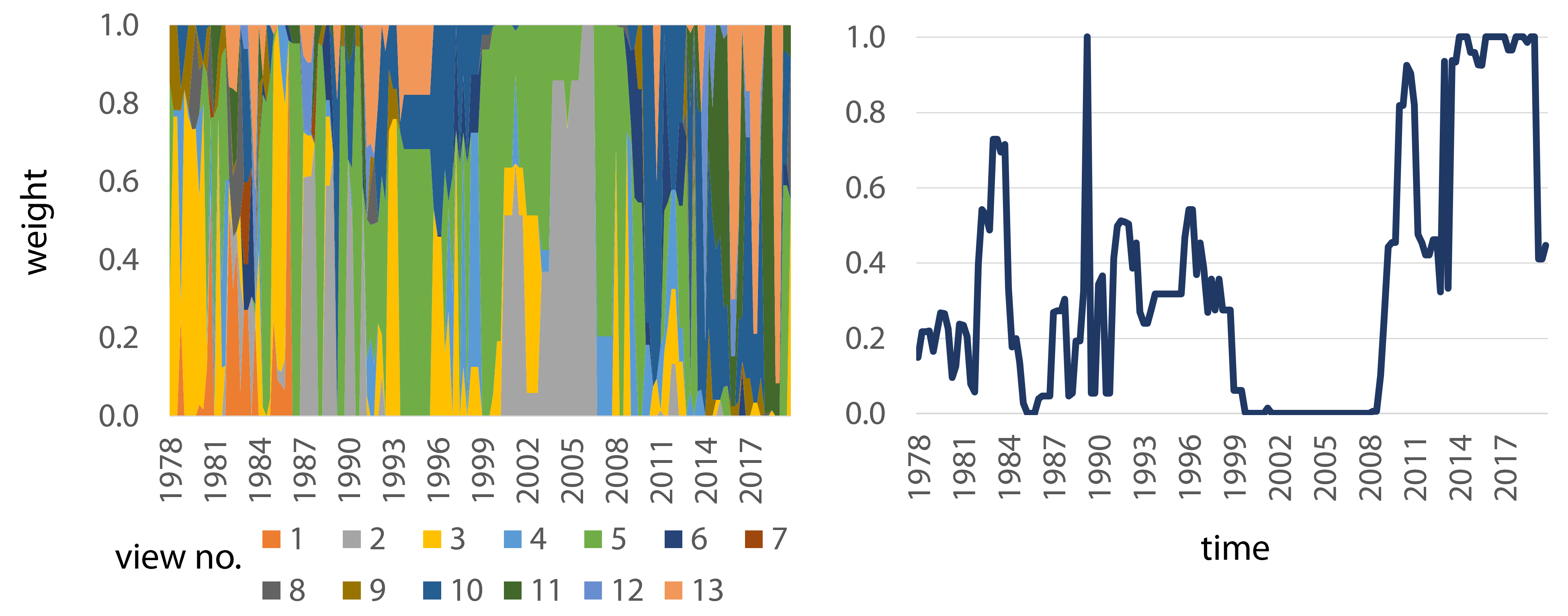}
	  	\subcaption*{\textit{Notes:} The area chart in the left panel shows the time-varying forecast combination weights for all the 13 views used to estimate the Markov-switching AR model. The chart goes from 1978Q1 to 2019Q4. The weights ($\mathbf{w}_2^{*}$) are obtained using the PIT-based optimization procedure described in the paper, where PIT stands for probability integral transform. The right panel plots the cumulative weight assigned to the views derived from Fed supervisory scenarios (views 6-13).  See Table \ref{tab:views} for the list of views.}
	  	\label{fig:w2}
	  \end{figure}

\begin{figure}[H]
	  	\caption{Optimal log-score-based posterior probabilities (prior $\boldsymbol{\pi}_1^{0*}$) over time}
	  	\centering
	  	\includegraphics[scale=0.15]{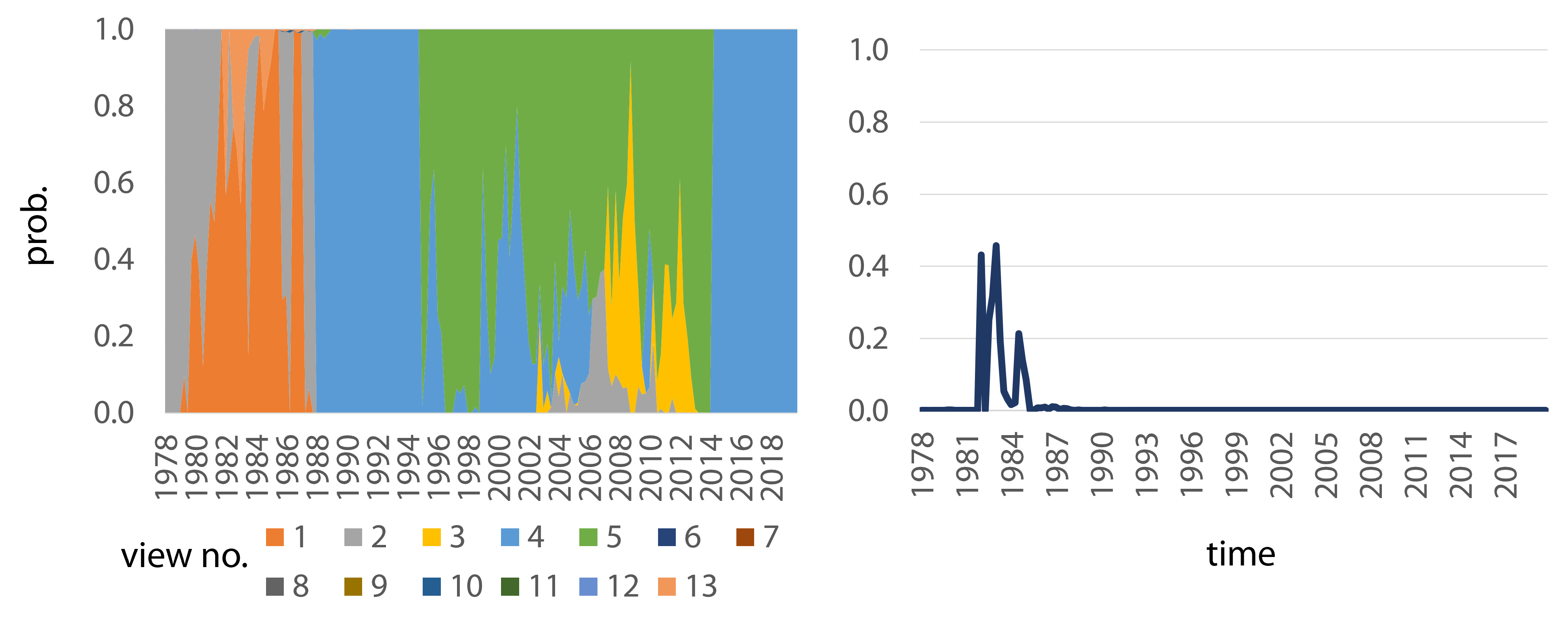}
	  	\subcaption*{\textit{Notes:} The area chart in the left panel shows the time-varying Bayesian posterior probabilities for all the 13 views used to estimate the Markov-switching AR model. The chart goes from 1978Q1 to 2019Q4. The underlying prior probabilities $\boldsymbol{\pi}_1^{0*}$ are obtained using the log-score-based optimization procedure described in the paper. The right panel plots the cumulative weight assigned to the views derived from Fed supervisory scenarios (views 6-13).  See Table \ref{tab:views} for the list of views.}
	  	\label{fig:pi1}
	  \end{figure}

\begin{figure}[H]
	  	\caption{Optimal PIT-based posterior probabilities (prior $\boldsymbol{\pi}_2^{0*}$) over time}
	  	\centering
	  	\includegraphics[scale=0.15]{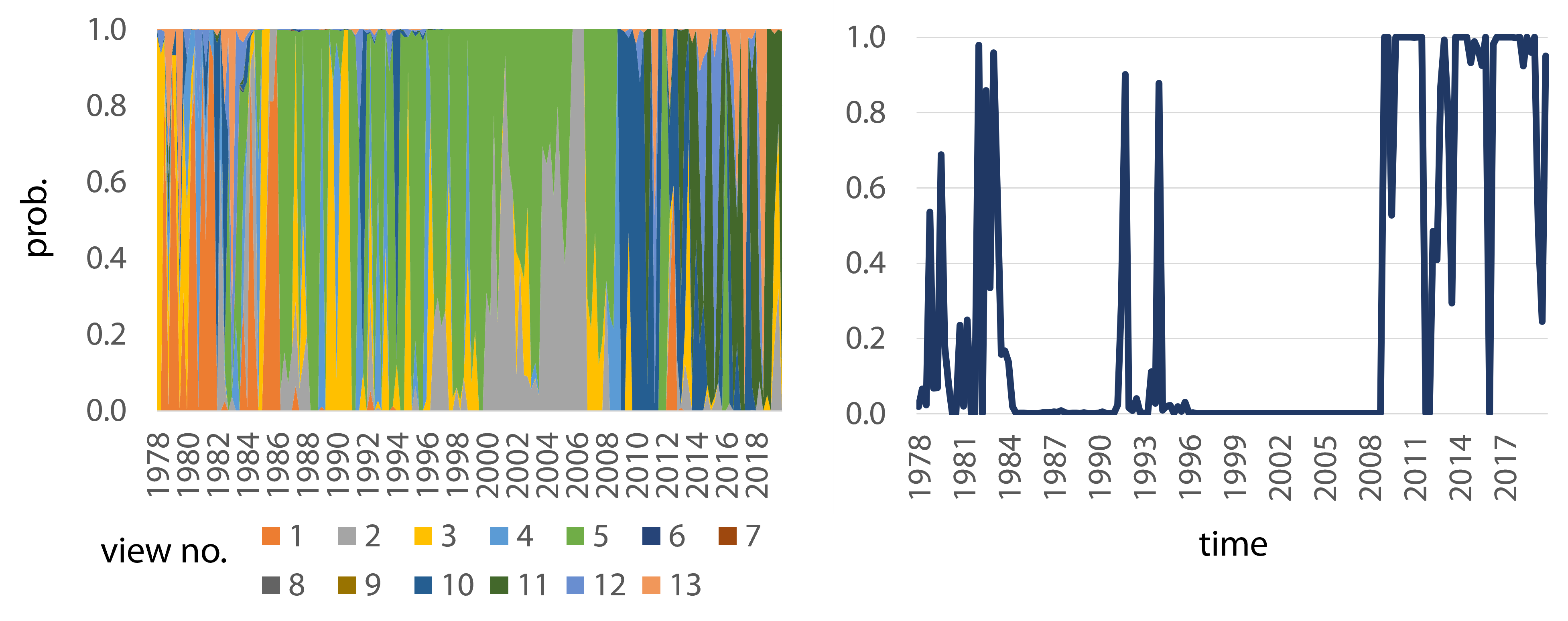}
	  	\subcaption*{\textit{Notes:} The area chart in the left panel shows the time-varying Bayesian posterior probabilities for all the 13 views used to estimate the Markov-switching AR model. The chart goes from 1978Q1 to 2019Q4. The underlying prior probabilities $\boldsymbol{\pi}_2^{0*}$ are obtained using the PIT-based optimization procedure described in the paper, where PIT stands for probability integral transform. The right panel plots the cumulative weight assigned to the views derived from Fed supervisory scenarios (views 6-13).  See Table \ref{tab:views} for the list of views.}
	  	\label{fig:pi2}
	  \end{figure}

       The approach can be used to evaluate the time-varying contribution of different views to the composite forecasts. Figures \ref{fig:w1}-\ref{fig:pi2} display the evolution over time of the optimal forecast weights and of the posterior probabilities resulting from the optimal priors. 
       In each figure, the area chart in the left panel shows the time-varying weights for all views from 1978Q1 to 2019Q4. The right panel plots the cumulative weight assigned to the views derived from the Fed supervisory scenarios.
       Figures \ref{fig:w1} and  \ref{fig:pi1} show the results of the optimization based on log scores, while Figures \ref{fig:w2} and \ref{fig:pi2} show the results of the optimization based on the PITs.
       As can be seen from Figures \ref{fig:w1} and \ref{fig:pi1}, the vague views tend to dominate in the case of log-score optimization.
        In terms of optimal weights $\mathbf{w}_1^{*}$, the cumulative weight of the Fed-based views lies in the range 7\%-42\% until 1990 and is zero afterwards. 
       As for optimized posteriors, the Fed-based views provide non-zero contributions only between 1982 and 1987. 
       Overall, the results indicate a minor role of Fed-based views in boosting density forecast accuracy.
 
        When the PIT-based optimization is considered, the contribution of the Fed-based views is much higher. 
        On average, they account for about 35\% of the combined forecasts in the case of optimal weights and 27\% in the case of posterior probabilities. 
        In particular, they dominate in the period following the Global Financial Crisis,  in terms of both $\mathbf{w}_2^{*}$ and $\boldsymbol{\pi}_{2}^*$.
        The also provide a substantial contribution in the early 1980s and during the 1990s.
       It is important to remark that using a single view is not sufficient to achieve well-calibrated forecasts. None of these views, when considered individually, leads to non-rejection of the PIT uniformity hypothesis in the KS test. Instead, the combination of different views is what drives the good results in terms of calibration.  
             
\begin{figure}[H]
	  	\caption{Density forecasts for 2016Q4}
	  	\centering
	  	\includegraphics[scale=0.5]{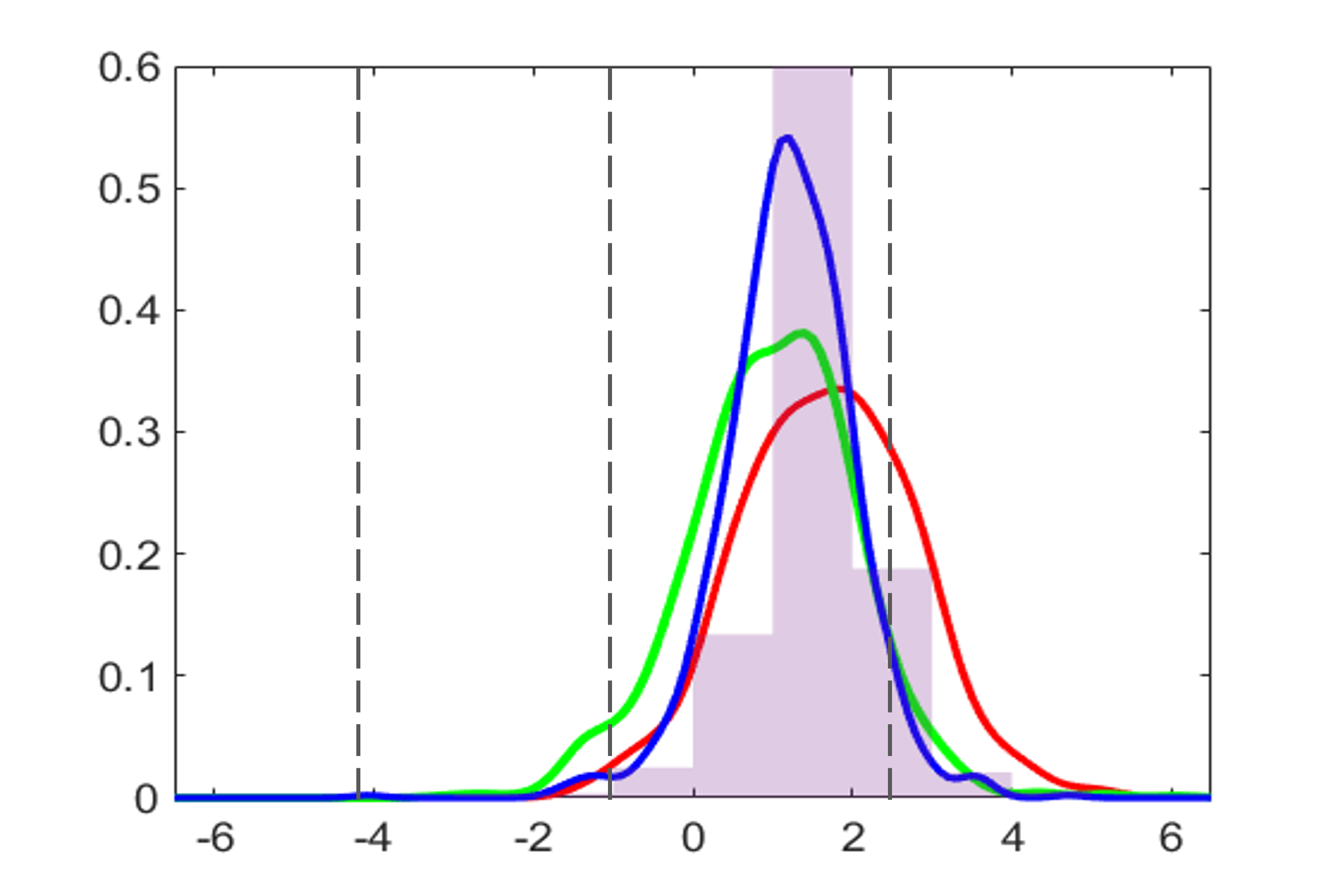}
	  	\subcaption*{\textit{Notes:} 
	  		The figure shows three probability density functions (PDFs): the density forecast of GDP growth from an AR model (red line), the density forecast generated by a 3-regime model using information from Fed stress tests scenarios (green line, view 9 in Table \ref{tab:views}), and the optimized scenario-augmented MSAR forecast (blue line), using combination weights $\mathbf{w}_{2}^{*}$.
	  		The three dashed vertical lines indicate the three scenarios for 2016Q4 contained in the 2016 Fed stress test.
	  		The figure also reports a histogram representing the distribution of the annual growth rate provided in 2016Q3 by the Survey of Professional Forecasters.
	  		The actual growth rate of GDP in 2016Q4 was 2 percent.
	  	}
	  	\label{fig:pdf_2016q3}
	  \end{figure}

       To illustrate the outcome of our procedure in more detail, Figure \ref{fig:pdf_2016q3} shows density forecasts for a specific quarter, 2016Q4, taken as an example. 
       The figure displays three probability density functions (PDFs): the density forecast from an AR model (red line), the forecast generated by a 3-regime MSAR model using a prior derived from the Fed stress test scenarios (green line; view 9 in Table \ref{tab:views}), and the optimized MSAR forecast (blue line), i.e., the final outcome of our procedure, using combination weights $\mathbf{w}_{2}^{*}$. 
       Three dashed vertical lines indicate the three discrete scenarios for 2016Q4 contained in the Fed's 2016 stress test.
       The figure provides an example of what can be obtained from our approach, compared to what is obtained from the single use of the two continuous/discrete approaches to forecast uncertainty.
       The (normal) density forecast of the AR model is a basic example of the classical density forecast outcome. 
       The predictive density from the 3-regime model is an example of scenario-augmented forecasts: it has a highly non-standard left-skewed profile, clearly reflecting a mixture of three different regime-specific normals, strongly influenced by the tight prior centered on external scenarios.
       The final (blue) forecast is an example of the ``optimal" combination of different densities (including the red and green ones reported in the figure). 
       It has an irregular and leptokurtic shape with two bumps in the tails, merging different forecasts as well as the prior information provided by external views.
        Finally, for comparison, the figure also reports a histogram representing the probability distribution of the annual growth rate provided in 2016Q3 (the forecast origin) by the Survey of Professional Forecasters.\footnote{
        	Note that the SPF does not provide a continuous density function of the quarterly GDP growth rate, but only probabilities associated with discrete intervals of annual GDP growth (e.g., the probability of that annual GDP growth in 2016 will be between 1\% and 2\%).
        }
       Our ``optimal" density forecasts are broadly in line with the SPF histogram, but of course allow for a more detailed characterization of the forecast distribution. 

\begin{figure}[H]
	  	\caption{Empirical distribution of the PITs}
	  	\centering
	  	\includegraphics[scale=0.5]{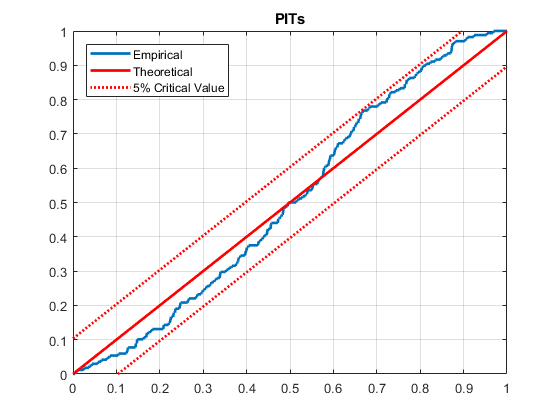}
	  	\includegraphics[scale=0.5]{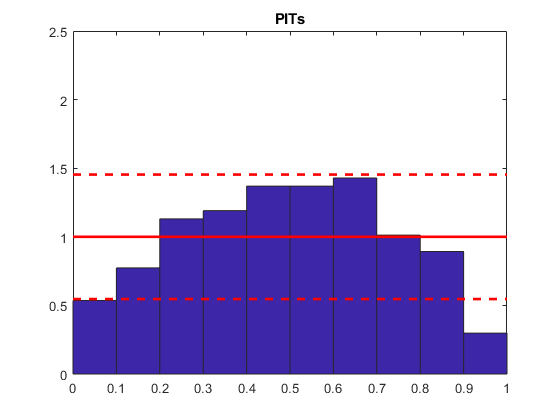}
	  	\subcaption*{\textit{Notes:} 
	  		The figure shows the empirical distribution of the probability integral transforms (PITs) associated with the optimized SA-MSAR density forecasts. The left panel shows the cumulative distribution function and the right panel shows the histograms of the PITs (normalized).
	  		In both panels, red solid lines indicate the uniform distribution and dashed lines indicate the 95\% confidence interval of the uniform distribution.}
	  	\label{fig:pit_tails}
	  \end{figure}

       Next, we examine the behavior of density forecasts across different parts of the GDP distribution. 
       To this aim, Figure \ref{fig:pit_tails} plots the entire empirical distribution of the PITs, along with the 95\% confidence intervals of the uniform distribution, as calculated by \citet{RossiSekhposyan2017}, which account for sample uncertainty.\footnote{For this analysis, we use Matlab code shared by \citet{RossiSekhposyan2017}. As mentioned by \citet{AdrianBoyarchenkoGiannone2019}, confidence bands should be taken as approximations when forecasts are computed using an expanding estimation windows.} 
       The figure shows both the cumulative distribution function (CDF) and a histogram of the PITs (normalized), with deciles along the horizontal axis.
       For perfectly calibrated forecasts, the empirical CDF of the PITs would lie on the 45-degree line, and the histogram bars would all have a height of 1. 
      	We report the empirical distribution of the PITs associated with the optimized SA-MSAR forecasts using combination weights $\mathbf{w}_{2}^{*}$.
       As the figure shows, the empirical CDF always lies within the 95\% confidence interval of the uniform distribution, in line with the result of the KS test in Table \ref{tab:results}. 
       In the histogram, all bars are within the confidence bands, except for the highest decile.
       This indicates that the density forecasts are well-behaved in most parts of the GDP growth distribution.
       Focusing on the tails,  we note that the bar at the lowest decile lies on the lower bound of the 95\% confidence interval, suggesting that the forecasts capture left tail risk quite adequately.
       The bar at the highest decile lies below the lower bound, indicating some over-dispersion of the forecasts in the upper tail.

      \subsubsection{Extension: applying density transformations}
      
      Lastly, we extend our approach by combining it with two types of density transformations that have been proposed in the literature to further improve density forecast performance:
      the beta-transformed linear pool (BLP; see \citealt{GneitingRanjan2010,GneitingRanjan2013})
      and the empirically-transformed linear opinion pools (ETLOP, \citealt{garrattetal2023}). 
      The BLP applies a transformation of the conditional forecast densities using the beta distribution. This can improve the calibration and, in particular, the dispersion of density forecasts produced by simple linear combinations.
      The ETLOP transforms the density forecasts so that it matches the 
      empirical cumulative distribution function of the target variable (GDP growth, in our case), in order to better accommodate non-Gaussianity.
      
      In the case of BLP, we select the values of the two parameters of the beta distribution using the same optimization scheme described before.
      In the case of ETLOP, at any time \textit{t} considered as the forecast origin, we use the empirical distribution of GDP growth up to time \textit{t}.
      
      Table \ref{tab:results_beta} shows the results. We report results using optimal weights, but very similar conclusions are obtained using optimal priors. 
      As a benchmark, we also report the results of BLP and ETLOP transformations applied to AR forecasts.
      As the table shows, the beta transformation further improves the density forecast performance of our scenario-augmented MSAR (SA-MSAR) forecasts, in terms of both log scores and PITs. The ADP is now 0.4-0.42 and the p-value of PIT uniformity test jumps to 0.68-0.81, depending on the specific  weights used.
      Also, our optimized SA-MSAR forecasts, once transformed by BLP, remain clearly superior to the benchmark BLP-transformed AR forecasts, in terms of both accuracy and calibration, although benchmark forecasts have also improved compared to their untransformed counterparts from Table \ref{tab:results}.
      Conversely, the ETLOP methodology does not appear to provide any improvement in this specific empirical application.
      
\begin{figure}[H]
        	\caption{Empirical distribution of the PITs: BLP-MSAR }
        	\centering
        	\includegraphics[scale=0.5]{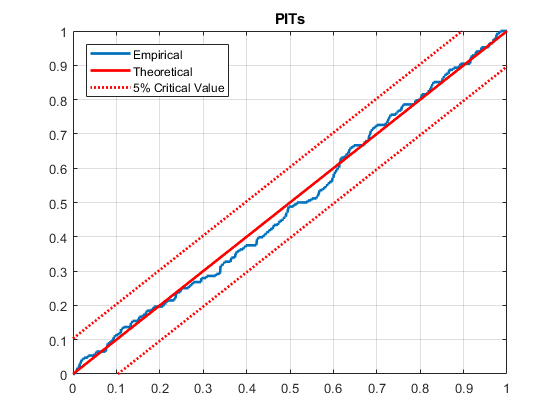}
        	\includegraphics[scale=0.5]{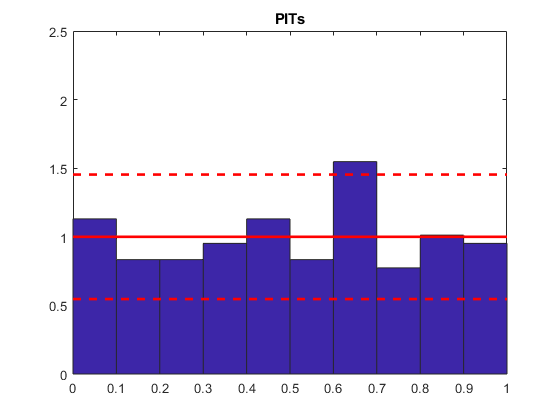}
        	\subcaption*{\textit{Notes:} 
        		The figure shows the empirical distribution of the probability integral transforms (PITs) associated with the optimized SA-MSAR density forecasts, transformed using a beta distribution (\citealt{GneitingRanjan2010,GneitingRanjan2013}). 
        		The left panel shows the cumulative distribution function and the right panel shows the histograms of the PITs (normalized).
        		In both panels, red solid lines indicate the uniform distribution and dashed lines indicate the 95\% confidence interval of the uniform distribution.
        	}	
        	\label{fig:pit_beta_tails}
        \end{figure}

	Finally, Figure \ref{fig:pit_beta_tails} shows the entire empirical distribution of the PITs in the case of beta-transformed optimal SA-MSAR forecasts (using combination weights $\mathbf{w}_2^{*}$). 
	In this case, also the highest decile lies within the 95\% confidence bands of the uniform distribution.

        \begin{table}[H]
        	\caption{Applying density transformations: BLP and ETLOP}
        	\centering
        	{
        		\begin{tabular}{l|ll|}
        			\hline \hline
        			forecasting method & APD   & KS     \\
        			\hline
        			ETLOP - AR 				& 0.21 	&	0.00	\\
        			BLP - AR             & 0.39 & 0.18 \\
        			\hline
        			ETLOP - SA-MSAR $\mathbf{w}_{1}^*$ & 0.26 & 0.00 \\
        			ETLOP - SA-MSAR $\mathbf{w}_{2}^*$ & 0.26 & 0.00\\
        			\textbf{BLP - SA-MSAR} $\mathbf{w}_{1}^*$  & \textbf{0.42} & \textbf{0.68} \\
        			\textbf{BLP - SA-MSAR} $\mathbf{w}_{2}^*$  & \textbf{0.40} & \textbf{0.81} \\
        			\hline \hline
        		\end{tabular}
        	}
        	\subcaption*{\textit{Notes:} 
        		The table reports the average predictive density (APD)
        		and the p-value of a Kolmogorov-Smirnov (KS) test of uniformity of the PITs associated with optimized scenario-augmented MSAR (SA-MSAR) forecasts, transformed using two alternative approaches: the beta-transformed linear pool (BLP; \citealt{GneitingRanjan2010,GneitingRanjan2013})
        		and the empirically-transformed linear opinion pools (ETLOP, \citealt{garrattetal2023}). 
        		As a benchmark, the table also reports the results for BLP- and ETLOP-transformed AR forecasts.}
        	\label{tab:results_beta}
        \end{table}

    \section{Conclusions}\label{sec:conclusions}

	We have developed an approach for generating density forecasts of macroeconomic variables using a variety of discrete economic scenarios provided by external sources.
	The approach is based on a Bayesian regime-switching model in which experts' views are translated into priors on economic regimes, and different views are pooled together to enhance density forecast performance.

	We have presented an empirical application in which density forecasts of U.S. GDP are formed using the supervisory scenarios defined by the Fed. 	
	We have shown that the approach is able to achieve both good forecast accuracy and correct calibration of predictive distributions, merging the flexibility of mixture predictive densities provided by regime-switching models with the benefits of forecast combination, which are well-established in the literature.

	Importantly, this methodology allows to evaluate the usefulness  of economists' views for density forecasting. 
	To illustrate this possibility, the empirical application tracks the contribution of Fed's scenarios to the optimized forecasts over time. 
	
	This approach appears particularly valuable in all contexts in which tail risk has a clear economic interpretation and when economic projections have to comply with external, possibly judgmental views.
	Researchers and practitioners interested in this kind of analysis may fine-tune the approach by tailoring the range of views to be considered and by selecting different objective functions in the optimization procedure.


     \pagebreak

    \bibliography{refs}

\begin{thebibliography}{44}
\newcommand{\enquote}[1]{``#1''}
\providecommand{\natexlab}[1]{#1}

\bibitem[{Aastveit et~al.(2018)Aastveit, Mitchell, Ravazzolo, and van
  Dijk}]{aastveitetal2018}
Aastveit, K.A., Mitchell, J., Ravazzolo, F., and van Dijk, H. (2018),
  \enquote{{The Evolution of Forecast Density Combinations in Economics},}
  Tinbergen Institute Discussion Papers 18-069/III, Tinbergen Institute.

\bibitem[{Acemoglu et~al.(2017)Acemoglu, Ozdaglar, and
  Tahbaz-Salehi}]{acemogluetal2017}
Acemoglu, D., Ozdaglar, A., and Tahbaz-Salehi, A. (2017),
  \enquote{Microeconomic Origins of Macroeconomic Tail Risks,} \emph{American
  Economic Review}, vol. 107~(1), pp. 54--108.

\bibitem[{Adrian et~al.(2020)Adrian, Morsink, and Schumacher}]{adrianIMF2020}
Adrian, M.T., Morsink, M.J., and Schumacher, M.L.B. (2020), \enquote{{Stress
  Testing at the IMF},} IMF Departmental Papers / Policy Papers 2020/001,
  International Monetary Fund.

\bibitem[{Adrian et~al.(2019)Adrian, Boyarchenko, and
  Giannone}]{AdrianBoyarchenkoGiannone2019}
Adrian, T., Boyarchenko, N., and Giannone, D. (2019), \enquote{Vulnerable
  Growth,} \emph{American Economic Review}, vol. 109~(4), pp. 1263--89.

\bibitem[{Alessandri and Mumtaz(2017)}]{alessandrimumtaz2017}
Alessandri, P. and Mumtaz, H. (2017), \enquote{Financial conditions and density
  forecasts for US output and inflation,} \emph{Review of Economic Dynamics},
  vol.~24, pp. 66--78.

\bibitem[{Ascari et~al.(2015)Ascari, Fagiolo, and
  Roventini}]{ascari_fagiolo_roventini_2015}
Ascari, G., Fagiolo, G., and Roventini, A. (2015), \enquote{Fat-tail
  distributions and business-cycle models,} \emph{Macroeconomic Dynamics},
  vol.~19~(2), p. 465–476.

\bibitem[{Bauwens et~al.(2017)Bauwens, Carpantier, and
  Dufays}]{bauwensetal2017}
Bauwens, L., Carpantier, J., and Dufays, A. (2017), \enquote{Autoregressive
  Moving Average Infinite Hidden Markov-Switching Models,} \emph{Journal of
  Business and Economic Statistics}, vol.~35~(2), pp. 162--182.

\bibitem[{Bidder and McKenna(2015)}]{BidderMcKenna2015}
Bidder, R. and McKenna, A. (2015), \enquote{{Robust stress testing},} Working
  Paper Series 2015-13, Federal Reserve Bank of San Francisco.

\bibitem[{Binder and Gross(2013)}]{bindergross103}
Binder, M. and Gross, M. (2013), \enquote{{Regime-switching global vector
  autoregressive models},} Working Paper Series 1569, European Central Bank.

\bibitem[{Borio et~al.(2014)Borio, Drehmann, and Tsatsaronis}]{borio2014}
Borio, C., Drehmann, M., and Tsatsaronis, K. (2014), \enquote{Stress-testing
  macro stress testing: Does it live up to expectations?} \emph{Journal of
  Financial Stability}, vol.~12, pp. 3--15.

\bibitem[{Chauvet and Potter(2013)}]{chauvetpotter2013}
Chauvet, M. and Potter, S. (2013), \enquote{Forecasting Output,} in G.~Elliott
  and A.~Timmermann, eds., \enquote{Handbook of Economic Forecasting,}
  \emph{Handbook of Economic Forecasting}, vol.~2, Elsevier, pp. 141--194.

\bibitem[{Clemen and Winkler(1999)}]{clemenwinkler1999}
Clemen, R.T. and Winkler, R.L. (1999), \enquote{Combining Probability
  Distributions From Experts in Risk Analysis,} \emph{Risk Analysis},
  vol.~19~(2), pp. 187--203.

\bibitem[{Cúrdia et~al.(2014)Cúrdia, Del~Negro, and
  Greenwald}]{curdiaetal2014}
Cúrdia, V., Del~Negro, M., and Greenwald, D.L. (2014), \enquote{Rare Shocks,
  Great Recessions,} \emph{Journal of Applied Econometrics}, vol.~29~(7), pp.
  1031--1052.

\bibitem[{Diebold et~al.(1998)Diebold, Gunther, and Tay}]{dieboldetal1998}
Diebold, F.X., Gunther, T., and Tay, A. (1998), \enquote{Evaluating density
  forecasts,} \emph{International Economic Review}, vol.~39~(4), pp. 863--883.

\bibitem[{Dueker(1997)}]{dueker1997}
Dueker, M.J. (1997), \enquote{{Strengthening the case for the yield curve as a
  predictor of U.S. recessions},} \emph{Federal Reserve Bank of St. Louis
  Review}, vol. Mar, pp. 41--51.

\bibitem[{Elliott and Timmermann(2016)}]{elliottandtimmermann2016}
Elliott, G. and Timmermann, A. (2016), \emph{{Economic Forecasting}}, Princeton
  University Press.

\bibitem[{Fagiolo et~al.(2008)Fagiolo, Napoletano, and
  Roventini}]{fagioloetal2008}
Fagiolo, G., Napoletano, M., and Roventini, A. (2008), \enquote{Are output
  growth-rate distributions fat-tailed? some evidence from OECD countries,}
  \emph{Journal of Applied Econometrics}, vol.~23~(5), pp. 639--669.

\bibitem[{Faust and Wright(2009)}]{FaustWright2009}
Faust, J. and Wright, J.H. (2009), \enquote{Comparing Greenbook and Reduced
  Form Forecasts Using a Large Realtime Dataset,} \emph{Journal of Business \&
  Economic Statistics}, vol.~27~(4), pp. 468--479.

\bibitem[{{Federal Reserve}(2013)}]{fed2017}
{Federal Reserve} (2013), \enquote{Policy Statement on the Scenario Design
  Framework for Stress Testing,} Federal register, 78, 71435–48.

\bibitem[{{Federal Reserve Board}(2018)}]{fed2018}
{Federal Reserve Board} (2018), \enquote{Supervisory Scenarios for Annual
  Stress Tests Required under the Dodd-Frank Act Stress Testing Rules and the
  Capital Plan Rule,} {Federal Reserve Board report}.

\bibitem[{Fr{\"u}hwirth-Schnatter(2006)}]{fruehwirthschnatter2006}
Fr{\"u}hwirth-Schnatter, S. (2006), \emph{{Finite Mixture and Markov Switching
  Models}}, Springer Series in Statistics, Springer, New York.

\bibitem[{Fr{\"u}hwirth-Schnatter(2008)}]{fruehwirthschnatter2008}
Fr{\"u}hwirth-Schnatter, S. (2008), \emph{Finite Mixture and Markov Switching
  Models. Implementation in MATLAB using the package bayesf Version 2.0},
  Springer Series in Statistics, Springer, New York.

\bibitem[{Ganics(2017)}]{ganics2017}
Ganics, G. (2017), \enquote{Optimal Density Forecast Combinations,} Working
  papers n. 1751, Banco de Espa{\~n}a.

\bibitem[{Garratt et~al.(2023)Garratt, Henckel, and Vahey}]{garrattetal2023}
Garratt, A., Henckel, T., and Vahey, S.P. (2023),
  \enquote{Empirically-transformed linear opinion pools,} \emph{International
  Journal of Forecasting}, vol.~39~(2), pp. 736--753.

\bibitem[{Genest and Zidek(1986)}]{GenestZidek1986}
Genest, C. and Zidek, J.V. (1986), \enquote{Combining Probability
  Distributions: A Critique and an Annotated Bibliography,} \emph{Statistical
  Science}, vol.~1~(1), pp. 114--135.

\bibitem[{Geweke and Amisano(2011)}]{gewekeandamisano2011}
Geweke, J. and Amisano, G. (2011), \enquote{Optimal prediction pools,}
  \emph{Journal of Econometrics}, vol. 164~(1), pp. 130--141.

\bibitem[{Gneiting and Ranjan(2010)}]{GneitingRanjan2010}
Gneiting, T. and Ranjan, R. (2010), \enquote{{Combining Probability
  Forecasts},} \emph{Journal of the Royal Statistical Society Series B:
  Statistical Methodology}, vol.~72~(1), pp. 71--91.

\bibitem[{Gneiting and Ranjan(2013)}]{GneitingRanjan2013}
Gneiting, T. and Ranjan, R. (2013), \enquote{{Combining predictive
  distributions},} \emph{Electronic Journal of Statistics}, vol.~7, pp. 1747 --
  1782.

\bibitem[{Gross et~al.(2022)Gross, Henry, and
  Rancoita}]{gross_henry_rancoita_2022}
Gross, M., Henry, J., and Rancoita, E. (2022), \emph{Macrofinancial Stress Test
  Scenario Design—for Banks and Beyond}, Cambridge University Press, p.
  77–97.

\bibitem[{Hall and Mitchell(2007)}]{hallandmitchell2007}
Hall, S.G. and Mitchell, J. (2007), \enquote{Combining density forecasts,}
  \emph{International Journal of Forecasting}, vol.~23~(1), pp. 1--13.

\bibitem[{Hamilton(2016)}]{hamilton2016}
Hamilton, J. (2016), \enquote{{Macroeconomic Regimes and Regime Shifts},} in
  J.B. Taylor and H.~Uhlig, eds., \enquote{{Handbook of Macroeconomics},}
  \emph{Handbook of Macroeconomics}, vol.~2, Elsevier, pp. 163--201.

\bibitem[{Hamilton(1989)}]{hamilton1989}
Hamilton, J.D. (1989), \enquote{A New Approach to the Economic Analysis of
  Nonstationary Time Series and the Business Cycle,} \emph{Econometrica},
  vol.~57~(2), pp. 357--384.

\bibitem[{Han and Leika(2019)}]{hanleika2019}
Han, F. and Leika, M. (2019), \enquote{{Integrating Solvency and Liquidity
  Stress Tests: The Use of Markov Regime-Switching Models},} IMF Working Papers
  2019/250, International Monetary Fund.

\bibitem[{Hansen(1994)}]{hansen1994}
Hansen, B.E. (1994), \enquote{Autoregressive Conditional Density Estimation,}
  \emph{International Economic Review}, vol.~35~(3), pp. 705--730.

\bibitem[{Jorion(2006)}]{jorion2006}
Jorion, P. (2006), \emph{{Value at Risk: The New Benchmark for Managing
  Financial Risk}}, McGraw Hill, New York, 3rd Edition.

\bibitem[{Krüger et~al.(2017)Krüger, Clark, and Ravazzolo}]{Kruegeretal2017}
Krüger, F., Clark, T.E., and Ravazzolo, F. (2017), \enquote{Using Entropic
  Tilting to Combine BVAR Forecasts With External Nowcasts,} \emph{Journal of
  Business \& Economic Statistics}, vol.~35~(3), pp. 470--485.

\bibitem[{{Moody's}(2018)}]{moodys2018}
{Moody's} (2018), \enquote{U.S. Macroeconomic Outlook Alternative Scenarios,}
  Tech. rep., Moody's Analytics.

\bibitem[{Pesaran et~al.(2006)Pesaran, Pettenuzzo, and
  Timmermann}]{pesaranetal2006}
Pesaran, M.H., Pettenuzzo, D., and Timmermann, A. (2006), \enquote{{Forecasting
  Time Series Subject to Multiple Structural Breaks},} \emph{The Review of
  Economic Studies}, vol.~73~(4), pp. 1057--1084.

\bibitem[{Robertson et~al.(2005)Robertson, Tallman, and
  Whiteman}]{Robertsonetal2005}
Robertson, J.C., Tallman, E.W., and Whiteman, C.H. (2005), \enquote{Forecasting
  Using Relative Entropy,} \emph{Journal of Money, Credit and Banking},
  vol.~37~(3), pp. 383--401.

\bibitem[{Rossi and Sekhposyan(2014)}]{RossiSekhposyan2014}
Rossi, B. and Sekhposyan, T. (2014), \enquote{Evaluating predictive densities
  of US output growth and inflation in a large macroeconomic data set,}
  \emph{International Journal of Forecasting}, vol.~30~(3), pp. 662--682.

\bibitem[{Rossi and Sekhposyan(2019)}]{RossiSekhposyan2017}
Rossi, B. and Sekhposyan, T. (2019), \enquote{Alternative tests for correct
  specification of conditional predictive densities,} \emph{Journal of
  Econometrics}, vol. 208~(2), pp. 638--657.

\bibitem[{Schorfheide and Song(2015)}]{SchorfheideSong1025}
Schorfheide, F. and Song, D. (2015), \enquote{Real-Time Forecasting With a
  Mixed-Frequency VAR,} \emph{Journal of Business \& Economic Statistics},
  vol.~33~(3), pp. 366--380.

\bibitem[{Wang et~al.(2022)Wang, Hyndman, Li, and Kang}]{wangetal2022}
Wang, X., Hyndman, R.J., Li, F., and Kang, Y. (2022), \enquote{Forecast
  combinations: An over 50-year review,} \emph{International Journal of
  Forecasting}.

\bibitem[{Wolters(2015)}]{Wolters2015}
Wolters, M.H. (2015), \enquote{Evaluating Point and Density Forecasts of DSGE
  Models,} \emph{Journal of Applied Econometrics}, vol.~30~(1), pp. 74--96.

\end{thebibliography}

	\pagebreak

	\begin{appendices}
	
	\section{Estimation and forecasting}
	
	\subsection{Bayesian estimation of MSAR with multiple views}
	
	Let us define here $\mathbf{y} = (y_1,\dots,y_T)$ and  $\mathbf{S} = (S_1,\dots,S_T)$. 
	Also, let 	$\boldsymbol{\vartheta} = (\beta_1,\dots,\beta_K,\sigma_1,\dots,\sigma_K,\alpha_1,\dots,\alpha_p,\boldsymbol{\xi})$ and 
	$\boldsymbol{\theta} = (\beta_1,\dots,\beta_K,\sigma_1,\dots,\sigma_K,\alpha_1,\dots,\alpha_p,)$.
	The posterior distribution 	$p(\boldsymbol{\vartheta}|\mathbf{y}) $ for model \eqref{eq:MSAR} is obtained using Bayes' theorem:
	\begin{equation}
		p(\boldsymbol{\vartheta}|\mathbf{y}) \propto p(\mathbf{y}|\boldsymbol{\vartheta}) p(\boldsymbol{\vartheta})
	\end{equation}
	where $p(\boldsymbol{\vartheta})$ is the prior on the parameters and $p(\mathbf{y}|\boldsymbol{\vartheta}) $ is the likelihood function, which in this case is a Markov mixture of normals (\citealt{fruehwirthschnatter2006}). Treating the state vector $\mathbf{S}$ as data, the Markov mixture likelihood can be expressed as the sum of the complete-data likelihood $p(\mathbf{y}, \mathbf{S} |\boldsymbol{\vartheta}) $ over all possible values of $\mathbf{S} $:
	\begin{align}
		p(\mathbf{y}|\boldsymbol{\vartheta}) &= \sum_{\mathbf{S} \in S_K} p(\mathbf{y}, \mathbf{S}|\boldsymbol{\vartheta}) \nonumber \\
		&= \sum_{\mathbf{S} \in S_K} p(\mathbf{y}|\mathbf{S},\boldsymbol{\theta}_{1}\, \dots, \boldsymbol{\theta}_K) p(\mathbf{S}|\boldsymbol{\xi}) \label{eq:likelihood}
	\end{align}  
	As shown in \citet{fruehwirthschnatter2006}, expression \eqref{eq:likelihood} factors in a convenient way that makes estimation easier. %
	In particular, if the prior assumes (i) the independence of the parameter vector $\boldsymbol{\theta}$  across regimes and (ii) the independence between parameters $\boldsymbol{\theta}$ and the transition matrix $\boldsymbol{\xi}$, i.e., 	 
	\begin{equation}
		p(\boldsymbol{\vartheta}) = \prod_{k=1}^{K} p(\boldsymbol{\theta}_k)p(\boldsymbol{\xi})
	\end{equation}
	then the complete-data posterior, i.e.,
	\begin{equation}
		p(\boldsymbol{\vartheta}|\mathbf{y},\mathbf{S}) \propto \prod_{k=1}^{K} p(\boldsymbol{\theta}_k|\mathbf{y},\mathbf{S}) p(\boldsymbol{\xi}|\mathbf{S})
	\end{equation}
	factors in the same way as the complete-data likelihood $p(\mathbf{y},\mathbf{S}|\boldsymbol{\vartheta}) $. This facilitates  the application of conventional Markov Chain Monte Carlo (MCMC) methods used for Bayesian estimation, in a context where, due to the Markov-switching behavior, the prior $p(\boldsymbol{\vartheta})$ and the posterior $p(\boldsymbol{\vartheta}|\mathbf{y})$ are not conjugate, and the posterior does not assume any convenient analytical form.  
	
	The posterior $p(\boldsymbol{\vartheta}|\mathbf{y})$  can be expressed as the sum of the posterior for the augmented parameter vector $(\mathbf{S},\boldsymbol{\vartheta})$ over all possible realizations of $\mathbf{S}$:
	\begin{align}
		p(\boldsymbol{\vartheta}|y) = \sum_{\mathbf{S} \in S_K} p(\mathbf{S},\boldsymbol{\vartheta}|\mathbf{y})
	\end{align} 
	In practice, Bayesian estimation samples from the joint posterior $  p(\mathbf{S},\boldsymbol{\vartheta}|\mathbf{y})$, using: 
	\begin{equation}
		p(\mathbf{S},\boldsymbol{\vartheta}|\mathbf{y}) \propto p(\mathbf{y}|\mathbf{S},\boldsymbol{\vartheta}) p(\mathbf{S}|\boldsymbol{\vartheta}) p(\boldsymbol{\vartheta})
	\end{equation}
	We estimate the model using MCMC methods and assuming independence priors of the following form:
	\begin{equation}
		p(\alpha_1,\dots,\alpha_p,, \beta_1, \dots, \beta_K, \sigma^2_1, \dots, \sigma^2_K) = \prod_{j=1}^{p} p(\alpha_j) \prod_{k=1}^{K} p(\beta_k) \prod_{k=1}^{K} p(\sigma^2_k) 
	\end{equation}
	If a Gamma hyper-prior is used for $C_0$, 
	 as is the case in our empirical application, 
	 the MSAR independence prior becomes:
	\begin{equation*}
		p(\alpha_1,\dots,\alpha_p, \beta_1, \dots, \beta_K, \sigma^2_1, \dots, \sigma^2_K, C_0) = \prod_{j=1}^{p} p(\alpha_j)  \prod_{k=1}^{K} p(\beta_k) \prod_{k=1}^{K} p(\sigma^2_k) p(C_0)
	\end{equation*}

	When several views are considered, Bayesian averaging across different views can performed in the following way.
	Let $\boldsymbol{\vartheta}_{K,i}^0$ denote the generic $i$-th view assuming $K$ states.
	First, the number of regimes is treated as uncertain. Accordingly, a discrete prior is defined for $K$, fixing a maximum number $\overline{K}$:
	\begin{gather}\label{eq:prior_K}
		\pi_K^0 = Pr\left(K\right) 
	\end{gather}
	for $K=1,\ldots,\overline{K}$, with $\sum_{K=1}^{\overline{K}} \pi_K^0 =1 $.
	Second, assuming that a number $P_K$ of alternative priors (views) are available for any given number of states $K$,
	a prior probability $\ \pi(\boldsymbol{\vartheta}_{K,i}^0|K)$ is assigned to $\boldsymbol{\vartheta}_{K,i}^0$, such that 
	$
	\sum_{i=1}^{P_K}{\pi(\boldsymbol{\vartheta}_{K,i}^0\left|K\right)}=1
	$.
	In other words, a discrete hierarchical prior is defined with respect to $\mathbf{\boldsymbol{\vartheta}}$. 
	The unconditional prior probability of $\boldsymbol{\vartheta}^0_{K,i}$ is equal to the joint prior probability of $\boldsymbol{\vartheta}^0_{K,i}$ and the number $K$ of regimes, i.e., $\pi(\boldsymbol{\vartheta}^0_{K,i}) = \pi(\boldsymbol{\vartheta}^0_{K,i},K)$. Using $ \pi^{0}_{K,i}$ to denote this unconditional probability, we have that:
	\begin{equation}\label{eq:pi0}
		\pi^{0}_{K,i}
		\equiv \pi(\boldsymbol{\vartheta}^0_{K,i}) =
		\pi(\mathbf{\boldsymbol{\vartheta}}_{K,i}^{0}\left|K\right)\pi_K^0
	\end{equation}

	Also, let $\boldsymbol{\pi}^{0}$ denote the vector of length $\sum_{K=1}^{\overline{K}}P_K\ $ containing the unconditional prior probabilities of all views, i.e., $\boldsymbol{\pi}^0= (\pi^0_{1,1},\dots,\pi^0_{\overline{K},P_{\overline{K}}} ) $.
	Finally, the posterior probabilities of the views depend on the prior vector $\boldsymbol{\pi}^0$, collecting the probabilities defined in \eqref{eq:pi0}, and on the marginal likelihood of the MSAR model under the different views. In particular, the posterior probability for view $\boldsymbol{\vartheta}^0_{K,i}$ is equal to the joint posterior probability of $\boldsymbol{\vartheta}^0_{K,i}$ and the number $K$ of regimes, i.e., $	\pi(\boldsymbol{\vartheta}^0_{K,i}|\mathbf{y}) = \pi(\boldsymbol{\vartheta}^0_{K,i},K|\mathbf{y})$, and is given by:
	\begin{equation}\label{eq:pi}
		\pi_{K,i}
		\equiv
		\pi(\boldsymbol{\vartheta}^0_{K,i}|\mathbf{y}) =\frac{p(\mathbf{y}|\boldsymbol{\vartheta}^0_{K,i}) \pi^{0}_{K,i}}{\sum_{K=1}^{\overline{K}}\sum_{j=1}^{P_K} p(\mathbf{y}|\boldsymbol{\vartheta}^0_{K,j}) \pi^{0}_{K,j} }
	\end{equation}
	where $ p(\mathbf{y}|\boldsymbol{\vartheta}^0_{K,i})=p(\mathbf{y}|\boldsymbol{\vartheta}^0_{K,i},K) = \int p(\mathbf{y}|\boldsymbol{\vartheta}_K,\boldsymbol{\vartheta}^0_{K,i},K) p(\boldsymbol{\vartheta}_K|\boldsymbol{\vartheta}^0_{K,i},K) d\boldsymbol{\vartheta}_K$, with $\boldsymbol{\vartheta}_K$ denoting the parameter vector in the MSAR model with $K$ regimes.

	\subsection{Computing density forecasts}
	
	Computing density forecasts from the MSAR model requires three steps. 
	Let $\mathbf{y}_t=(y_0,y_1,\dots,y_t)$. Also, let us assume that the current time period is $T$ and the forecast horizon is one period. 
	The first step consists in using the MCMC algorithm to sample both the current unobserved regime $S_T$ and the MSAR parameters $\boldsymbol{\vartheta}$ from the posterior distribution $p(\mathbf{S},\boldsymbol{\vartheta}|\mathbf{y}_T)$. Let $(\boldsymbol{\vartheta}^{(d)}, S^{(d)}_T)$ denote a generic MCMC draw.
	Next, each draw is used to forecast the future state of the economy. Taking $S_T^{(d)}$ as the starting value, a stochastic forecast $S_{T+1}^{(d)}$ is computed using the matrix of transition probabilities $\boldsymbol{\boldsymbol{\xi}}^{(d)}$, i.e., based on \eqref{eq:transition_probability}.
	Third, $y_{T+1}^{(d)}$ is sampled from the normal predictive density $p(y_{T+1}, | \mathbf{y}_T,\boldsymbol{\vartheta}^{(d)}, S_{T+1}^{(d)}) $. In particular, 
	\begin{equation}
		y_{T+1}|\mathbf{y}_T,\boldsymbol{\vartheta}^{(d)}, S_{T+1}^{(d)}=k
		\sim 
		\mathcal{N} \left( 
		\sum_{j=1}^{p}\alpha^{(d)}_j y_{T+1-j} + \beta_k^{(d)} ,\sigma^{(d)2}_k
		\right)
	\end{equation}
	
	Conditional on knowing the state of the economy in the future period $T+1$, the predictive distribution of $y_{T+1}$ is a Normal for any given parameter vector. However, since the future state of the economy is unknown, the density forecast of $y_{T+1}$ produced by the MSAR will be a mixture of the different regime-specific normals, where the mixture weights are given by the probabilities of the economy ending up in the different possible regimes at $T+1$. As a result, the MSAR generally produces non-normal forecast distributions. Also, the predictive densities are non-linear in $y_{T}$ and heteroskedastic (\citealt{fruehwirthschnatter2006}). 
	More specifically, assuming a known number of regimes $K$ and a known parameter vector $\mathbf{\boldsymbol{\vartheta}}$, the one-step-ahead density forecast at time $T$ is the following finite mixture of $K$ normal components: 
	\begin{equation}
		p\left(y_{T+1}|\mathbf{y}_{T},\mathbf{\boldsymbol{\vartheta}}\right)=\ \sum_{k=1}^{K}p\left(y_{T+1}|\mathbf{y}_{T},\boldsymbol{\theta}_k\right)Pr\left(S_{T+1}=k|\mathbf{y}_{T},\mathbf{\boldsymbol{\vartheta}}\right)
	\end{equation}
	Next, as a result of Bayesian estimation, the density forecast for any given view integrates out parameter uncertainty: 
	\begin{equation}
		p\left(y_{T+1}|\mathbf{y}_{T},\boldsymbol{\vartheta}_{K,i}^0 \right)= 
		\int
		p\left(y_{T+1}|\mathbf{y}_{T},\boldsymbol{\vartheta}_K,\boldsymbol{\vartheta}_{K,i}^0 \right)
		p(\boldsymbol{\vartheta}_K|\mathbf{y}_{T},\boldsymbol{\vartheta}_{K,i}^0) d\boldsymbol{\vartheta}_K
	\end{equation}          
	where $\boldsymbol{\vartheta}_K$, as before, denotes the parameter vector when $K$ regimes are assumed.

	\subsection{Prior on the regime-switching variance in the empirical application}
	Based on the properties of the Gamma and inverted Gamma distributions (see, e.g., \citealt{fruehwirthschnatter2006}), it holds that:
	\begin{align}
	\text{E}(\sigma^2_k|C_0) &= \frac{C_0}{c_0-1}\\
	\text{Var}(\sigma^2_k|C_0)&=\frac{C_0^2}{(c_0-1)^2(c_0-2)}\\
	\text{E}(C_0) &= \frac{g_0}{G_0} \\ 
	\text{Var}(C_0) &= \frac{g_0}{G_0^2} \\
	\text{E}(C_0^2) &= \left(\frac{g_0}{G_0}\right)^2  + \frac{g_0}{G_0^2} \\ 
	\end{align}
	
	Given the hyperparameter values $c_0=3$, $g_0=0.5$ and $G_0=0.5$, it follows that:
	\begin{align}
	\text{E}(\sigma^2_k) &= \frac{\text{E}(C_0)}{c_0-1} = 0.5\\
	\text{Var}(\sigma^2_k) &= \text{E}(\text{Var}(\sigma^2_k|C_0)) + \text{Var}(\text{E}(\sigma^2_k|C_0)) =\\
	&= \frac{\text{E}(C_0^2)}{(c_0-1)^2(c_0-2)} + \frac{\text{Var}(C_0^2)}{(c_0-1)^2}  = \\[0.5em]
	&= \frac{3}{4} + \frac{1}{2}=  1.25
	\end{align}


	\section{Fed supervisory scenarios}

\begin{table}[H]
	\caption{Fed stress tests 2015-2018: scenarios of GDP growth}
	\centering
		\scalebox{0.9}[0.9]{
	\begin{tabular}{l|lll|lll|lll|lll|}
		\hline \hline
		& \multicolumn{3}{c|}{2015} & \multicolumn{3}{c|}{2016}& \multicolumn{3}{c|}{2017}& \multicolumn{3}{c|}{2018}\\
		time	& base& adv.&sev.&base& adv.&sev.&base& adv.&sev. &base& adv.&sev.\\
		\hline
		2014Q4 & 3   & -0.6 & -3.9 &     &      &      &     &      &      &     &      &      \\
		2015Q1 & 2.9 & -1.3 & -6.1 &     &      &      &     &      &      &     &      &      \\
		2015Q2 & 2.9 & -0.2 & -3.9 &     &      &      &     &      &      &     &      &      \\
		2015Q3 & 2.9 & 0.2  & -3.2 &     &      &      &     &      &      &     &      &      \\
		2015Q4 & 2.9 & 0.3  & -1.5 &     &      &      &     &      &      &     &      &      \\
		2016Q1 & 2.9 & 0.8  & 1.2  & 2.5 & -1.5 & -5.1 &     &      &      &     &      &      \\
		2016Q2 & 2.9 & 1.2  & 1.2  & 2.6 & -2.8 & -7.5 &     &      &      &     &      &      \\
		2016Q3 & 2.9 & 1.7  & 3    & 2.6 & -2   & -5.9 &     &      &      &     &      &      \\
		2016Q4 & 2.9 & 1.8  & 3    & 2.5 & -1.1 & -4.2 &     &      &      &     &      &      \\
		2017Q1 & 2.7 & 1.8  & 3.9  & 2.4 & 0    & -2.2 & 2.2 & -1.5 & -5.1 &     &      &      \\
		2019Q4	& 2.7 & 1.9  & 3.9  & 2.5 & 1.3  & 0.4  & 2.3 & -2.8 & -7.5 &     &      &      \\
		2017Q3 & 2.6 & 2    & 3.9  & 2.3 & 1.7  & 1.3  & 2.4 & -2   & -5.9 &     &      &      \\
		2017Q4 & 2.6 & 2.2  & 3.9  & 2.3 & 2.6  & 3    & 2.3 & -1.5 & -5.1 &     &      &      \\
		2018Q1 &     &      &      & 2.6 & 2.6  & 3    & 2.4 & -0.5 & -3   & 2.5 & -1.3 & -4.7 \\
		2018Q2 &     &      &      & 2.4 & 3    & 3.9  & 2.4 & 1    & 0    & 2.8 & -3.5 & -8.9 \\
		2018Q3 &     &      &      & 2.3 & 3    & 3.9  & 2.4 & 1.4  & 0.7  & 2.6 & -2.4 & -6.8 \\
		2018Q4 &     &      &      & 2.3 & 3    & 3.9  & 2.3 & 2.6  & 3    & 2.5 & -1.3 & -4.7 \\
		2019Q1 &     &      &      & 2.1 & 3    & 3.9  & 2   & 2.6  & 3    & 2.3 & -0.7 & -3.6 \\
		2019Q2 &     &      &      &     &      &      & 2.1 & 3    & 3.9  & 2.3 & 0.4  & -1.3 \\
		2019Q3 &     &      &      &     &      &      & 2.1 & 3    & 3.9  & 2.1 & 1    & -0.2 \\
		2019Q4 &     &      &      &     &      &      & 2   & 3    & 3.9  & 2   & 2.5  & 2.8  \\
		2020Q1 &     &      &      &     &      &      & 2   & 3    & 3.9  & 2.1 & 2.8  & 3.5  \\
		2020Q2 &     &      &      &     &      &      &     &      &      & 2.1 & 3    & 4    \\
		2020Q3 &     &      &      &     &      &      &     &      &      & 2.1 & 3.2  & 4.2  \\
		2020Q4 &     &      &      &     &      &      &     &      &      & 2.1 & 3.3  & 4.5  \\
		2021Q1 &     &      &      &     &      &      &     &      &      & 2.1 & 3.3  & 4.5 \\
		\hline \hline
	\end{tabular}
}
	\subcaption*{\textit{Notes:}  For each year between 2015 and 2018 the table reports the baseline, adverse and severely adverse supervisory scenarios for U.S. GDP growth (annualized quarter-on-quarter, in percentage) included in the annual stress test conducted by the Federal Reserve (see Federal Reserve Board 2014, 2016, 2017, 2018).}
	\label{tab:fed_scenarios}
\end{table}

\end{appendices}

\end{document}